\documentclass[letter,11pt]{article}%
\usepackage{amsmath,bm}
\usepackage{amsfonts}
\usepackage{amsthm}
\usepackage{amssymb}

\usepackage{comment}
\usepackage{lscape}
\usepackage{multibib}
\usepackage{float}
\usepackage{graphicx}
\usepackage{rotating}
\usepackage[]{subfigure}
\usepackage[height=9.5in,left=1in,right=0.75in,bottom=1in]{geometry}
\usepackage[colorlinks=true,citecolor=blue, linkcolor=black]{hyperref}
\usepackage{natbib}
\usepackage{color}
\usepackage{colortbl}
\usepackage{appendix}
\usepackage{paralist}
\usepackage{threeparttable}
\usepackage{multirow}
\usepackage{amsmath}%
\usepackage{array}
\newcolumntype{P}[1]{>{\centering\arraybackslash}p{#1}}

\providecommand{\U}[1]{\protect\rule{.1in}{.1in}}
\usepackage{titlecaps}
\Addlcwords{for a an and with in of to the for}

\RequirePackage{hypernat}

\RequirePackage{hypernat}

\theoremstyle{plain}

\newtheorem{theorem}{Theorem}

\newtheorem{assumption}{Assumption}

\newtheorem{lemma}[theorem]{Lemma}

\newtheorem{proposition}[theorem]{Proposition}

\newtheorem{remark}{Remark}

\catcode`~=11
\newcommand{\urltilde}{\kern -.15em\lower .7ex\hbox{~}\kern .04em}
\catcode`~=13
\def \@seccntformat#1{\csname the#1\endcsname.\quad}
\makeatother
\numberwithin{equation}{section}
\hypersetup{pdftitle={}, pdfsubject={}, pdfauthor={}, pdfkeywords={} }

\begin{document}
	\title{Machine Learning Debiasing with Conditional Moment Restrictions: An Application to LATE\thanks{An earlier version of this paper circulated under the title:
			\textquotedblleft On the Existence and Information of Orthogonal Moments for
			Inference\textquotedblright, arXiv 2303.11418. We thank Guido Imbens and participants at many
			institutions for useful comments. Research funded by Ministerio de Ciencia e
			Innovaci\'{o}n grant PID2021-127794NB-I00 and Comunidad de Madrid, grants
			EPUC3M11 (VPRICIT) and H2019/HUM-589.}}
	\author{Facundo Arga\~{n}araz\thanks{Department of Economics. E-mail:
			\href{mailto:farganar@eco.uc3m.es}{farganar@eco.uc3m.es}. Website: \href{https://argafacu.github.io}{https://argafacu.github.io}.}\\\textit{Universidad Carlos III de Madrid}
		\and Juan Carlos Escanciano\thanks{Department of Economics. E-mail: \href{mailto:jescanci@indiana.edu}{jescanci@econ.uc3m.es}. Website: \href{https://sites.google.com/view/juancarlosescanciano}{https://sites.google.com/view/juancarlosescanciano}.}\\\textit{Universidad Carlos III de Madrid}}
	\date{\today}
	\maketitle
	
	\begin{abstract}
		{Models with Conditional Moment Restrictions (CMRs) are popular in economics. These models involve finite and infinite dimensional parameters. The infinite dimensional components include conditional expectations, conditional choice probabilities, or policy functions, which might be flexibly estimated using Machine Learning tools. This paper presents a characterization of locally debiased moments for regular models defined by general semiparametric CMRs with possibly different conditioning variables. These moments are appealing as they are known to be less affected by first-step bias. Additionally, we study their existence and relevance. Such results apply to a broad class of smooth functionals of finite and infinite dimensional parameters that do not necessarily appear in the CMRs. As a leading application of our theory, we characterize debiased machine learning for settings of treatment effects with endogeneity, giving necessary and sufficient conditions. We present a large class of relevant debiased moments in this context. We then propose the Compliance Machine Learning Estimator (CML), based on a practically convenient orthogonal relevant moment. We show that the resulting estimand can be written as a convex combination of conditional local average treatment effects (LATE). Altogether, CML enjoys three appealing properties in the LATE framework: (1) local robustness to first-stage estimation, (2) an estimand that can be identified under a minimal relevance condition, and (3) a meaningful causal interpretation. Our numerical experimentation shows satisfactory relative performance of such an estimator. Finally, we revisit the Oregon Health Insurance Experiment, analyzed by \cite{finkelstein2012oregon}. We find that the use of machine learning and CML suggest larger positive effects on health care utilization than previously determined.}  \vspace{1mm}
		
		\begin{description}
			\item[Keywords:] Local Treatment Effects; Debiased Inference; Machine Learning.
			
			\item[\emph{JEL classification:}] C14; C31; C36.\newpage
			
		\end{description}
	\end{abstract}
	\newgeometry{height=9in,left=1in,right=0.75in,bottom=1in}
	
	\section{Introduction}
	\label{intro}
	
	Models with Conditional Moment Restrictions (CMRs) are popular in economics and statistics, appearing in regressions, quantile models, discrete choice models, demand estimation, and missing data problems, among others; see \cite{chen2016methods} for a wide range of settings with CMRs. An important subclass subsumes semiparametric specifications, where the model depends on infinite and finite dimensional parameters. The finite dimensional parameters might be some coefficient or treatment effect, which are the focus of the analysis.  The infinite dimensional components include conditional expectations, conditional choice probabilities, or policy functions, which might be flexibly estimated using Machine Learning tools, such as Lasso, Random Forest, Boosting, Neural Networks, and the like.   
	
	These tools have been proved useful in undertaking flexible estimation of high-dimensional objects. They are able to deliver reliable predictions by achieving a suitable bias-variance trade-off. Thus, the estimation of these high-dimensional first-step parameters is necessarily biased. Such bias might carry over into the estimation of the second-step finite-dimensional parameters. Typically, this second-step bias decays at a rate slower than the $\sqrt{n}$ rate, invalidating standard inference. In particular, the resulting estimator might not be $\sqrt{n}-$consistent and asymptotically normal, as shown in \cite{chernozhukov2022locally} (CEINR). In practical terms, this issue translates into confidence intervals with poor coverage and test statistics with sizes different from the nominal level.   
	
	This paper presents a characterization of Locally Robust (LR)/Debiased/Orthogonal moments for regular models defined by general semiparametric CMRs with possibly different conditioning variables. These moments are appealing as they are known to be less affected by first-stage bias; see \cite{chernozhukov2018double}. As such, they can restore standard inference on the parameters of interest, delivering estimators that are $\sqrt{n}-$consistent and asymptotically normal. Additionally, apart from presenting a general construction of orthogonal moments, we study their existence and relevance.

	Our theoretical results complement those in \cite{arganaraz2023existence} (AE), who have studied the existence, relevance, and construction of debiased moments for general smooth functionals in semiparametric likelihood settings. Instead, in this paper, we deal with CMRs models, with residual functions depending on finite and infinite dimensional parameters. As in AE, we deliver moments for general smooth functionals, which might depend on all the parameters indexing the model, regardless of its dimension. The smooth functionals do not need to be an explicit argument of the residual functions appearing in the CMRs. Examples include average structural functions, counterfactual effects and other downstream inferences.
	
	Our results build on those in \cite{chen2018overidentification}. These authors have previously characterized the regularity and the tangent space for the class of models we consider, which are important ingredients in our construction. However, they have not investigated orthogonal moments, as we do. Thus, our results complement those of \cite{chen2018overidentification}. 
	
	The construction of LR moments in our setting is conceptually straightforward. It is based on the use of Orthogonal Instrumental Variables (OR-IVs), which were introduced in Section 7 of the first version of CEINR, \cite{chernozhukovetaldebiased2016}. These OR-IVs are transformations of the conditioning variables that are orthogonal to all pathwise derivatives of the CMRs with respect to the nuisance parameters. However, not all OR-IVs give informative inferences, in the sense of being able to detect Pitman local alternatives from the true parameter value. The OR-IVs that deliver informative orthogonal moments for the parameter of interest are denoted as ORthogonal Relevant IVs (ORR-IVs). We provide a characterization of ORR-IVs. To the best of our knowledge, the literature has no discussion of ORR-IVs, a concept we find useful.

	Our findings complement previous developments in the literature on debiasing machine learning. \cite{chernozhukovetaldebiased2016} provided a general characterization of LR moments and OR-IVs for the finite-dimensional parameters of the CMRs---also allowing for different conditioning variables, including the optimal OR-IV. \cite{chernozhukov2018double} considered the special case of a common set of conditioning variables. Differently from these references, we consider as parameters of interest smooth functionals that do not necessarily appear in the CMRs. Furthermore, we show that not all OR-IVs are informative and characterize those that give informative inferences (ORR-IVs).

	A leading application of our general theory is the treatment effect setting where selection to treatment is endogenous. For this problem, a model that has been very popular in practice is the partially linear model with endogeneity (PLME). Methods relying on IVs are known to be useful for learning treatment effects in experiments where treatment randomization is not possible (e.g., when treatment participation is contingent on subjects' decisions), but there exists a source of exogenous variation in the treatment induced by an instrument; see \cite{imbensangrist94}, \cite{imbensangristrubin96}, \cite{abadie2003semiparametric}, \cite{kolesa2013estimation}, \cite{sloczynski2020should}, \cite{huntington2020instruments}, \cite{abadie2024instrumental}, and \cite{mogstad2024instrumental} for a recent review. 
	
	Within this setting, we characterize the set of moments that are debiased for the treatment coefficient, giving a simple necessary condition under which this set is not-empty (existence). In this set, any orthogonal moment is indexed by a particular IV choice, which is a function---with finite second moment---of pre-treatment exogenous characteristics and an IV that is excluded from the main equation, $X$ and $Z_1$, respectively. From an orthogonality standpoint, all such choices might be employed, as all of them deliver a debiased moment. A feature of these orthogonal IVs (OR-IVs) is that they are the result of recentering an initial IV, in the sense that their conditional expectation, given the vector of pre-treatment characteristics, is zero. The benefits of recentering IVs is not new; see \cite{borusyak2020non}, for example, who analyze bias issues from non-random exposure to exogeneous variation.
	
	For each IV choice, a particular estimand can be defined. As with any other IV, a relevance condition must hold to identify the estimand, which might or not be satisfied. We characterize a class of ORR-IVs in this context that satisfy the relevance condition, provided that $Z_1$ is a predictor of the treatment, given $X$. We then select, based on implementation considerations, a convenient ORR-IV and propose the \textit{Compliance Machine Learning} (CML) estimator. The estimand of this estimator can be seen as resulting from a 2SLS where the treatment is instrumented by the recentered fitted values from the first-stage, i.e., $\mathbb{E}\left[D|Z_1,X\right]$, where $D$ is treatment. With treatment and $Z_1$ binary, we show that the resulting estimand, i.e., the probability limit of CML, has a meaningful nonparametric interpretation as it can be written as a weighted average of conditional local average treatment effects (LATE) under misspecification, with non-negative weights. Altogether,  CML enjoys three appealing properties in the LATE framework: (1) local robustness to first-stage estimation, (2) an estimand that can be identified under a minimal relevance condition, and (3) a meaningful causal interpretation.

	Our analysis is closely related to \cite{robustestimationcaetanoetal2023}, who first studied the identification of weighted averages of conditional LATEs with non-negative weights within regression discontinuity designs (RDD) with covariates. Since Fuzzy RDD can be seen as a binary IV at the threshold, their results apply to our IV context. Moreover, \cite{robustestimationcaetanoetal2023} characterize the IV that ``maximizes" the relevance condition, important for identification and inference. These authors do not exploit machine learning tools for estimating first steps, as we do, and thus we acknowledge the key orthogonality property that our moments, based on OR-IVs, enjoy.
	
	It is well known that when both $D$ and $Z_1$ are binary, in the presence of covariates, LATE is a convex combination of conditional LATEs. Moreover, under the typical LATE framework \citep[][]{imbensangrist94, angrist1995two,  abadie2003semiparametric}, the estimand of an IV estimator is not necessarily a convex combination of conditional LATEs, and thus it does not have a meaningful nonparametric interpretation. As \cite{sloczynski2020should} showed,  the typical IV estimand, which uses $Z_1$ as the IV, does not deliver a nonparametric causal interpretation, unless a strong monotonicity condition is imposed; see also \cite{kolesa2013estimation}. The problem is that the typical choice $Z_1$ leads to an estimand that is constructed as a weighted sum of conditional LATEs, but those weights might take negative values, when there are compliers but not defiers at some values of the covariates and there are defiers but not compliers elsewhere.\footnote{Following the terminology of \cite{imbensangristrubin96}, ``compliers" are those subjects that are treated only when induced by the instrument $Z_1$, and ``defiers" are those who are treated only when not induced by $Z_1$.} This is a weaker version of the typical monotonicity assumption that rules out defiers in the entire population. \cite{angrist1995two} (and \cite{angrist2009mostly}) study an estimand that presents this desirable causal interpretation, however, it has not been extensively used in practice.\footnote{See footnote 1 of \cite{sloczynski2020should}.} An important ingredient of this estimator is controlling nonparametrically for covairates, which has been shown a necessary condition for obtaining TSLS  estimands with interpretable weights by \cite{blandhol2022tsls}. More specifically, \cite{angrist1995two}'s proposal requires running a fully saturated model, using as instruments $Z_1$ and the interaction terms between $Z_1$ and indicator variables for each possible value of $X$. This is equivalent to running a separate first stage for each possible combination of covariate values. When the dimension (and the support) of $X$ is large, the task of implementing such an estimator becomes formidable.  
	
	Conveniently, we show that the probability limit of CML, which always controls for covariates nonparametrically, coincides with that  of \cite{angrist1995two}, and thus it can be written as a convex combination of conditional LATEs, where the weights are guaranteed to be non-negative, under the weaker version of monotonicity and misspecification. However, CML can be implemented straightforwardly using machine learning tools and cross-fitting without having to run saturated specifications.\footnote{Cross-fitting has been extensively used in the semiparametric literature; see, e.g., \cite{bickel1982adaptive}, \cite{klaassen1987consistent}, \cite{vandervaart98}, \cite{robins2008higher}, \cite{Zheng2010AsymptoticTF}, \cite{chernozhukov2018double}.} 
	
	Our proposed estimator is not new. Independently, \cite{chernozhukov2018double}, in their footnote 8, suggests using a LR moment based on our recommended ORR-IV, the same moment that we derived. Our main point of departure is that we provide theoretical and practical justifications to support the view that our specific ORR-IV shall be used in applied work, over other choices. \cite{chernozhukov2018double} also recommends a different OR-IV which leads to the Double Debiased Machine Learning estimator (DML). They recommend the use of the recentered $Z_1$. Our theory and Monte Carlo simulations suggest that CML, using the recentered $\mathbb{E}\left[D|Z_1,X\right]$, might outperform DML. Altogether, this paper makes the case that CML is a useful empirical tool.
	
	The nonparametric interpretation of our preferred estimand has been acknowledged by previous works. \cite{kolesa2013estimation} proposes a (leave-one-out) unbiased jackknife instrumental variable estimator \citep[][]{phillips1977bias,angrist1999jackknife} whose estimand coincides with that in \cite{angrist1995two}. Nevertheless, the moment that both \cite{angrist1995two} and \cite{kolesa2013estimation}  consider is not LR. Our recommended ORR-IV  choice, which leads to our estimand with a nonparametric causal interpretation, coincides with the optimal (or efficient) augmented IV proposed by \cite{coussens2021improving}, under the assumption that $Z_1$ and $X$ are independent and with homoskedastic residuals in the structural equation. In this case, our recommended ORR-IV captures the strength of the instrument (i.e., the effect of $Z_1$ on $D$, when $X$ is controlled for nonparametrically), which is related to the probability of compliance with the instrument. \cite{coussens2021improving} argue that  their IV (and thus our preferred ORR-IV) can be interpreted as being weighting observations according to such probability, similarly to \cite{abadie2024instrumental} who propose an IV estimator that implicitly weights groups of observations according to the strength of the instrument in the first stage.\footnote{As observed by the authors, with a heterogeneous second-stage, \cite{abadie2024instrumental}'s weighting scheme has the same causal interpretation as \cite{kolesa2013estimation}'s.}  \cite{coussens2021improving} recommend the use of machine learning to obtain the instrument as we do, but we do not assume independence between the instrument and covariates.

	We showcase the utility of the CML by revisiting the Oregon Health Insurance Experiment, analyzed by \cite{finkelstein2012oregon}. In this experiment, it is well-known that the treatment, an indicator of health insurance coverage, is potentially endogenous, as compliance was not perfect. Nevertheless, there exists a source of exogenous variation in the treatment that can be exploited.  By considering a low-dimensional setting, \cite{finkelstein2012oregon} find considerable and statistically significant increases in the number of prescription drugs and outpatient visits in health centers, one year after treatment took place. Instead, we utilize additional pre-treatment variables and two different machine learning tools. With this, CML reports qualitatively the same effect on drug prescriptions and outpatient visits, i.e., health insurance access has a positive effect on these. Nevertheless, CML suggests larger effects. This is an important aspect since increased health care utilization would be translated into larger health costs from the supply side.\footnote{Additional use of health care services might bring improved health status as well; however, a one-year span might not be large enough to detect these positive effects.} The recommended estimator indicates that the treatment increases the number of drug prescriptions by roughly 0.4 units (std err. = 0.17), which represents a 14\% larger effect relative to that in \cite{finkelstein2012oregon}. That effect increases up to 0.63/0.507 units (std err.=0.208/0.152) if an equivalent implementation algorithm is employed. Moreover, it would suggest that health insurance coverage most likely increases the number of outpatient visits by  1.3 units (std err. = 0.172/0.190), which amounts to a 20\% larger effect than that in \cite{finkelstein2012oregon}. Additionally, machine learning supports the finding that a positive impact on emergency room visits is observed, on extensive and intensive margins, a result that could not be uncovered by \cite{finkelstein2012oregon}. The use of a given machine learning algorithm does not seem to influence our findings. What is more, we point out that CML, in most cases, reports more precise estimates than other competing alternatives, including DML. We believe that these empirical results and additional Monte Carlo exercises provide pieces of evidence that support the use of CML in applied work.

	The rest of the paper is organized as follows. Section \ref{generalsetting} presents our general setting with CMRs and varying conditioning variables. Section \ref{existencegral} provides a characterization of orthogonal moments for such settings, and studies their existence and relevance. Section \ref{PLME} applies the previous general theory to the PLME. In particular, Section \ref{omexistence} gives a general construction of debiased moments in this example and studies their existence. A subclass of ORR-IV for this setting is introduced in Section \ref{srelevance}. Section \ref{sinterpretation} shows that a practically convenient member of such subclass leads to an estimand that can be written as a convex combination of conditional LATEs, under a weaker version of monotonicity and misspecification of the model. Section \ref{sestimation} describes the CML estimator and its implementation through machine learning tools and cross-fitting. A Monte Carlo experimentation that compares CML with competing alternatives and illustrates our theory is shown in Section \ref{smontecarlo}. Our empirical application is outlined in Section \ref{soregon}. Finally, Section \ref{sconclusion} provides final remarks. An Appendix contains additional theoretical results, the asymptotic theory of the introduced estimator, and some details on our Monte Carlo experiments.

	\bigskip \noindent \textbf{Notation:} The norm $\left|\left|\cdot\right|\right|$ is a generic norm. Let $\mathbb{P}$ be a probability distribution. Let $L^2$ be the space of functions of $Z$ that are square-integrable when $ Z \sim \mathbb{P}$, where the precise meaning of $Z$ will be established below.\footnote{Technically, we should index $L^2$ by some $\sigma-$finite measure. We avoid this to simplify our notation.} In addition, let $L^2_0$ be a subset of $L^2$ with the additional mean-zero restriction. Similar definitions apply for objects such as $L^2(V)$ and $L^2_0(V)$ for functions of $V$, an arbitrary random variable. For any arbitrary subset $\mathcal{K}$, let $\overline{\mathcal{K}}$ denote the closure of $\mathcal{K}$ and $\overline{\mathcal{K}}^{\perp}$ be its orthocomplement, when a topology and inner product $\left<\cdot,\cdot\right>$ is defined. Moreover, $\Pi_{\overline{\mathcal{K}}}$ is the orthogonal projection operator onto $\overline{\mathcal{K}}$. Let $J < \infty$ and  $V = \left(V_1, V_2, \cdots, V_J\right)^{'}$, we say that a vector-valued function $f(V) = \left(f_1(V_1), \cdots, f_J(V_J)\right)^{'}$ is in $L^2(V) \equiv \bigotimes^J_{j=1} L^2(V_j)$ when each of the elements in a such vector belongs to the corresponding $L^2\left(V_j\right)$, $j=1,\cdots,J$.

	\section{General Setting}
	\label{generalsetting}
	
	We observe a vector $Z = \left(Y,X,W\right)$ from a distribution $\mathbb{P}_0$ that belongs to a semiparametric model  
	\begin{equation}
		\mathcal{P}=\{\mathbb{P}:\mathbb{E}_{\mathbb{P}}\left[  \left.  \rho
		_{j}(Y,\theta_{0},\eta_{0})\right\vert W_{j}\right]  =0\text{ a.s. for all
		}j=1,...,J,\text{ }\theta_{0}\in\Theta,\text{ }\eta_{0}\in \Xi\}, \label{CMRmodel}%
	\end{equation}
	where $\Theta\subset\mathbb{R}^{p}$ and $ \eta_{0} \in \Xi$ is a vector of real-valued measurable  functions of $X$ that might depend on additional unknown parameters that we do not specify, and $\Xi$ is some linear vector space. Specifically, for a generic $ \eta \in \Xi$,  $\eta = \left(\eta_1,\cdots,\eta_{d_\eta}\right)$ with $\eta_s \equiv \eta_s\left(X\right)$. The residual functions $\rho_j$ are known up to the parameters $\theta_0$ and $\eta_0$, with finite second moment. $\rho_j$ is allowed to depend on the entire function $\eta_0$, not only on its evaluation at a particular realization of $X$. Observe that in the case where $W_j$ is a constant, for some $j$, we have an unconditional moment. Moreover, note that we are not imposing anything regarding how the conditioning variables relate. They might or might not have elements in common. When $Z \sim \mathbb{P}_0$, we let $\mathbb{E}_{\mathbb{P}_0}\left[\cdot\right] \equiv \mathbb{E}\left[\cdot\right]$.
	
	Model \eqref{CMRmodel} is remarkably general. It encompasses a wide range of settings previously studied by the semiparametric literature. It covers the model considered by \cite{ai2012semiparametric} with nested CMRs. This generalizes the parametric setting in \cite{chamberlain1987asymptotic} with one conditioning variable, as well as the semiparametric settings in \cite{chamberlain1992efficiency} and \cite{ai2003efficient} with one conditioning variable. It also generalizes the parametric models in \cite{chamberlain1992cond} and \cite{brown1998efficient} with nested CMRs; see also \cite{ai2007estimation}. Moreover, our model \eqref{CMRmodel}  also addresses the setting studied by \cite{ackerberg2014asymptotic}, which has semiparametric CMRs with possibly non-nested or overlapping conditioning sets. More broadly, a nonparametric version of our model has been considered by \cite{chen2018overidentification}, who provide conditions for the regularity of the model and have characterized its tangent space.\footnote{By regularity, we mean that the tangent space of the full model is linear and a subset of $L^2$.}   
	
	\section{Existence, Relevance, and Construction of LR moments}
	\label{existencegral}
	Let $\psi_0 = \psi\left(\lambda_{0}\right) \in \mathbb{R}^{d_\psi}$, where $\lambda_{0} = \left(\theta_{0},\eta_{0}\right) \in \Lambda$, be a parameter of interest. A leading example is $\psi(\lambda_0) = \theta_0$, but we allow for more general functionals as well. In particular, the residuals $\rho_j$ are not necessarily functions of $\psi\left(\lambda_{0}\right)$. Our goal is to conduct inference on $\psi\left(\lambda_{0}\right)$ locally robust to the unknown $\lambda_0$. More specifically, we aim to base inference for $\psi_0$ on the moment
	$$
	\mathbb{E}\left[g\left(Z,\lambda_0, \kappa_0\right)\right] = 0,
	$$
	where
	$$
	g(Z,\theta_{0},\eta_{0}, \kappa_0)= \sum^J_{j=1}\rho_{j}(Z,\theta_{0},\eta_{0})\kappa_{0j}(W_{j}), 		
	$$
	and $\kappa_0 \in L^2(W)$ is a vector of IV-choices, i.e, a $J$-dimensional vector of real-valued functions $\kappa_{0j}$ of the conditioning variables $W_{j}$ such that each of its entries has finite second moments. Convenient choices of $\kappa_0$ will be provided below. 
	
	We are interested in inference on smooth functionals. They are defined as follows. Let $\dot{\psi}(h) = d\psi(\lambda_\tau)/d\tau$, where $d\lambda_\tau/d\tau = \left(\delta,b\right)\equiv h$ represents the local change in $\lambda_\tau = \left(\theta_\tau, \eta_\tau\right)$ from a departure of the true $\lambda_0$, and where henceforth, $\frac{d}{d\tau}$ denotes the derivative from the right (i.e., from non-negatives values of $\tau$), evaluated at $\tau = 0$. We assume throughout that $h \in \overline{\Delta(\lambda_0)}$, for a suitable closed space 
	$\overline{\Delta(\lambda_0)}$  and that $\dot{\psi}$ is a continuous function when we see it as a mapping on $\overline{\Delta(\lambda_0)}$. Let $\overline{\Delta(\lambda_0)} \subseteq \mathbf{H} = \mathcal{H}_\theta \times \mathcal{H}_\eta$, for Hilbert spaces $\mathcal{H}_\theta$ and $\mathcal{H}_\eta$, and where $\mathbf{H}$ is a Hilbert space with inner product $\left<h_1,h_2\right>_{\mathbf{H}} = \left<\delta_1,\delta_2\right>_{\mathcal{H}_\theta} + \left<b_1,b_2\right>_{\mathcal{H}_\eta}$. Linearity and continuity of $\dot{\psi}$ implies that there exists a unique $r_{\psi} = \left(r_{\psi,t}\right)^{d_\psi}_{t=1}$ with $r_{\psi,t} \in \overline{\Delta(\lambda_0)}$, $t=1,\cdots,d_\psi$, such that for all $h \in \overline{\Delta(\lambda_0)}$ 
	$$
	\dot{\psi}(h) = \left<h,r_{\psi}\right>_{\mathbf{H}}.
	$$
	In the case of a multi-dimensional functional $\psi(\lambda_0)$, the previous representation is understood componentwise. In the literature, $r_{\psi}$ is known as the Riesz representer, whose existence and uniqueness follows by the Riesz representation theorem. Ifa debiased momoment exists, we construct it under the null hypothesis assuming that $\psi(\lambda_0)$ is fixed at a known $\psi_0$, which can be seen as a \textit{restriction} to our model. In this restricted model, it must be the case that $\psi(\lambda_\tau) = \psi_0$ for all small $\tau$ implies
	$\dot{\psi}(h) = \left<h,r_{\psi}\right>_{\mathbf{H}} = 0$, where $\lambda_\tau \in \Lambda$ and $\lambda_\tau \neq \lambda_0$. This observation is key in our definition of orthogonality of moments with respect to nuisanse components. A debiased moment is less sensitive to first-stage estimation, or more formally,
	\begin{align}
		\frac{d}{d\tau} \mathbb{E}\left[g\left(Z,\lambda_\tau, \kappa_0\right)\right] & = 0, \;\;\; \text{for all}\; h \in \overline{\Delta\left(\lambda_0\right)} \;\;\;\text{such that}\;\;\;\left<h,r_{\psi}\right>_{\mathbf{H}} = 0, \label{cond1} \\ 
		\mathbb{E}\left[g\left(Z,\lambda_0, \kappa\right)\right] & = 0, \;\;\; \text{for all}\;\; \kappa \in L^2\left(W\right) \label{cond2}.
	\end{align}
	Notice that \eqref{cond1} implies that the Gateaux derivative of a moment based on $g$ is zero at the truth, which is an important property to construct a moment locally insensitive to first-stage bias. Observe that, for reasons that will become clear shortly, we introduce a new nuisance parameter: $\kappa_0$. This is the key component to achieve local robustness of a moment based on $g$. Condition \eqref{cond2} establishes that the moment is \textit{globally} robust to this parameter.

	Next, we provide a general construction for a $g$ that satisfies \eqref{cond1}-\eqref{cond2}. Define
	\[
	m_{j}(W_{j},\theta,\eta):=\mathbb{E}\left[  \left.  \rho
	_{j}(Y,\theta,\eta)\right\vert W_{j}\right]  \text{ a.s.}%
	\]
	and the Fréchet derivatives $\nabla m(W,\theta,\eta)[h]=(\nabla m_{1}(W_{1}%
	,\theta,\eta)[h],...,\nabla m_{J}(W_{J},\theta,\eta)[h]),$ where
	\[
	\nabla m_{j}(W_{j},\theta,\eta)[h]=\frac{d}{d\tau}m_{j}(W_{j},\theta
	+\tau\delta,\eta+\tau b),\text{ }h=(\delta,b)\in \overline{\Delta(\lambda_0)},
	\]
	and note that $\theta_\tau \equiv \theta + \tau \delta$, $\eta_\tau \equiv \eta + \tau b$. Define next 
	\[
	\nabla m:h\in\overline{\Delta(\lambda_0)}\rightarrow\nabla m(W,\theta,\eta)[h]\in%
	{\textstyle\bigotimes\nolimits_{j=1}^{J}}
	L^2(W_{j}) \equiv L^2\left(W\right).
	\]
	Note that $\nabla m$ is a linear and continuous operator, mapping from $\overline{\Delta(\lambda_0)}$ into $ L^2\left(W\right)$. Let $ L^2\left(W\right)$ be endowed with the inner product $\left<f_1,f_2\right>_{L^2\left(W\right)} = \sum^J_{j=1}\mathbb{E}\left[f_{1j}\left(W_j\right) f_{2j}\left(W_j\right)\right]$, where $f_{s} = \left(f_{s1}, \cdots, f_{sJ}\right)^{\prime}$, $s=1,2$, and thus $L^2(W)$ is a Hilbert space. Therefore, $\nabla m$ is a linear operator from a Hilbert space into a Hilbert space and we denote its adjoint operator by $\left(\nabla m\right)^{*}$. The adjoint operator, by definition, satisfies for any $f \in L^2(W)$
	$$
	\left<\nabla m[h], f\right>_{L^2(W)} =  \left<h, \left(\nabla m\right)^{*}f\right>_{\mathbf{H}}.
	$$
	Furthermore, the range of $\nabla m$, evaluated at $\lambda = \lambda_0$, when restricted to the set of $h$ such that $\left<h,r_{\psi}\right>_{\mathbf{H}} = 0$ is defined as 
	$$
	\mathcal{V}_{r^{\perp}_\psi}: = \left\{f \in L^2(W): f = \nabla m\left(W,\theta_0,\eta_0\right)[h], \;\;\; h \in \overline{\Delta(\lambda_0)}\;\;\; \text{and}\;\;\;  \left<h,r_{\psi}\right>_{\mathbf{H}} = 0\right\}.
	$$

	The following result characterizes the entire class of LR moments for model \eqref{CMRmodel}, i.e., if a function belongs to this class is a LR moment and any LR moment belongs to this class. Additionally, it provides necessary and sufficient conditions for the existence of those functions, i.e., for a non-empty class of debiased moments. \cite{chernozhukovetaldebiased2016} first established this result for the special case where $\psi(\lambda_0) = \theta_0$, but more generally did not discuss existence or relevance.

	\begin{theorem}
		\label{ThCMR} Let Assumptions 4.1 and 4.2 in \cite{chen2018overidentification} hold.
		Then, the class of LR moments for \eqref{CMRmodel} is composed of functions $g\in L_{0}^{2}$  of the form
		$$
		g(Z,\theta_{0},\eta_{0}, \kappa_0)= \sum^J_{j=1}\rho_{j}(Z,\theta_{0},\eta_{0})\kappa_{0j}(W_{j}), 		
		$$
		where 
		$$
		\sum^J_{j=1}\mathbb{E}\left[\nabla m_{j}(W_{j},\theta_0,\eta_0)[h]\kappa_{0j}(W_{j})\right] = 0,
		$$
		for all $\left<h,r_{\psi}\right>_{\mathbf{H}} = 0$. Moreover, there exist no LR moments for $\psi(\lambda_{0})$ if $\overline{\mathcal{V}}_{r^{\perp}_\psi} = L^2(W)$, i.e., $\mathcal{V}_{r^{\perp}_\psi}$ is dense in $L^2(W)$ . Furthermore, any $0 \neq \kappa_0$ leading to a LR moment satisfies $\left(\nabla m\right)^{*}\kappa_0 = C^{\prime}r_\psi$ for a vector $C \in \mathbb{R}^{d_\psi}$.
	\end{theorem}
	Assumptions 4.1 and 4.2 in \cite{chen2018overidentification} are sufficient conditions
	for regularity of the model (\ref{CMRmodel}). We refer to all the IVs such that $\left(\nabla m\right)^{*}\kappa_0 = C^{\prime}r_\psi$ as ``Orthogonal Instrumental Variables" (OR-IVs) as they are transformations of the conditioning variables leading to orthogonal moments. 
	We are also able to determine when orthogonal moments are informative/relevant. We say that a LR moment is informative for $\psi(\lambda_0)$ if
	$$
	\frac{d}{d\tau} \mathbb{E}\left[g\left(Z,\lambda_\tau,\kappa_0\right)\right] = \sum^J_{j=1}\mathbb{E}\left[\nabla m_{j}(W_{j},\theta_0,\eta_0)[h]\kappa_{0j}(W_{j})\right] \neq 0,
	$$
	for some $h \in \overline{\Delta(\lambda_0)}$ with $\left<h,r_\psi\right>_\mathbf{H} \neq 0$, i.e., when $\psi(\lambda_\tau)$ is not restricted to be fixed at $\psi_0$. In other words, an informative debiased moment, leads to tests with non-trivial asymptotic local power for testing the null hypothesis $H_0: \psi(\lambda_0) =  \psi_0 $ against some local alternatives that converge to $H_0$ at the parametric rate $\sqrt{n}$.  
	To characterize Orthogonal and Relevant IVs (ORR-IVs), we define
	$\nabla m_{\psi}:=\left(  \nabla m\right)  [r_\psi],\ $where again the
	application of $\nabla m$ is componentwise when $\psi(\cdot)$ is multivariate, and the collection of all OR-IVs
	$$
	\mathcal{D} : = \left\{f \in L^2\left(W\right): \left(\nabla m\right)^{*}f= C^{\prime}r_\psi\;\;\; \text{for a vector}\;\;\; C \in \mathbb{R}^{d_\psi}\right\}.
	$$
	
	\begin{theorem}
		\label{informative2} There exists an ORR-IV for
		$\psi(\lambda_{0})$ iff $\nabla m_{\psi}$ is not orthogonal to $ \mathcal{D}$. Moreover, in that case, an ORR-IV is given by the orthogonal projection of $\nabla m_{\psi}$ onto $\mathcal{D}$. \bigskip
	\end{theorem}
	
	In Theorem \ref{ThCMR} and Theorem \ref{informative2}, we deal with conditional
	moment restrictions. However, in some applications researchers are only
	willing to assume orthogonality restrictions on linear subspaces of $L^2%
	(W)$, such as in high-dimensional linear regressions. We generalize the
	previous results to this case in the following remark.
	
	\begin{remark}
		\label{orthogMR} The results of this section are extended to general
		orthogonality restrictions as follows. Suppose the model is defined by%
		\[
		\mathcal{P}=\{\mathbb{P}:\mathbb{E}_{\mathbb{P}}\left[  \rho_{j}%
		(Y,X,\theta_{0},\eta_{0})\varphi_{j}\left(  W_{j}\right)  \right]  =0\text{
			for all }\varphi_{j}\in\Gamma_{j}\subseteq L^2(W_{j}),\text{ all
		}j=1,...,J\},
		\]
		where $\theta_{0}\in\Theta,$ $\eta_{0}\in\Xi,$ and $\Gamma_{j}$ is a closed
		linear subspace of $L^2(W_{j}).$ Then, all the previous results on existence
		and relevance hold with $\nabla m_{j}$ replaced by $\Pi_{\Gamma_{j}}%
		\circ\nabla m_{j}$ and $\left(  \nabla m_{j}\right)  ^{\ast}$ replaced by
		$\left(  \nabla m_{j}\right)  ^{\ast}\circ\Pi_{\Gamma_{j}}.$ Likewise, the
		definition of $\nabla m_{\psi}$ is now $\nabla m_{\psi}=\Pi_{\Gamma}%
		\circ\left(  \nabla m\right)  [r_\psi],$ where $\Pi_{\Gamma}$ is applied
		coordinatewise, i.e. $\Pi_{\Gamma}\circ\left(  \nabla m\right)  =(\Pi
		_{\Gamma_{j}}\circ\left(  \nabla m_{j}\right)  )_{j=1}^{J}.$\bigskip
	\end{remark}

	\section{Partially Linear Model with Endogeneity}
	\label{PLME}
	We shall now apply the general results of Section \ref{existencegral} to the Partially Linear Model with Endogeneity. Consider the situation where we are interested in learning the effect of some treatment $D$ (e.g., health care insurance coverage) on some outcome $Y$ (e.g., first care services utilization). Treatment is endogenous, potentially due to lack of compliance, but we have access to a vector of variables $Z_1$, correlated with the treatment and only affecting the outcome through $D$. In addition, we observe a vector of pre-treatment characteristics $X$. We observed $n$ independent and identically distributed copies of $Z = \left(Y_{1},D,X,Z_{1}\right)$, $Z_i = \left(Y_{i},D_i,X_i,Z_{1i}\right)$, $i = 1,\cdots,n$. The outcome $Y$ is assumed to follow the model\footnote{For simplicity of notation, we now let $Y$ denote the outcome variable only.}
	\begin{equation*}
		Y = \theta_0 D + \phi(X) + \varepsilon, \;\;\; \mathbb{E}\left[\varepsilon|Z_1,X\right] = 0,
	\end{equation*}
	where $\phi$ is an unknown function of covariates that we are not interested in and is assumed to be an element of $L^2(X)$.\footnote{In what follows, conditional moment restrictions are always assumed to be satisfied a.s. We omit this condition to simplify our notation.} Note that $\phi$ can be identified as $\phi(X) = \mathbb{E}\left[\left.Y - \theta_0D\right|X\right]$, and thus we shall write 
	\begin{equation}
		\label{eqmodel}
		\tilde{Y} = \theta_0 \tilde{D} + \varepsilon, \;\;\; \mathbb{E}\left[\varepsilon|Z_1,X\right] = 0,
	\end{equation}
	where $\tilde{V} = V - \mathbb{E}\left[\left. V \right|X\right]$, for a generic random variable $V$.\footnote{The version of the model obtained by partialing out covariates $X$ is equivalent to the one in terms of $\phi$. The advantage of model \eqref{eqmodel} is that we can express it in terms of conditional expectations of observables, objects that can easily be estimated in practice with a wide range of machine learning tools.} This is a version of the standard partially linear model \citep[][]{robinson1988root}.

	\subsection{Orthogonal Moments and Existence}
	\label{omexistence}
	Note that  \eqref{eqmodel} leads to the conditional moment restriction 
	\begin{equation}
		\label{cmr}
		\mathbb{E}\left[\left.\tilde{Y} - \theta_0\tilde{D}\right|Z_1,X\right] = 0.
	\end{equation}
	In terms of our notation above,  the conditioning variables are collected in $W = \left(Z_1,X\right)$, and $\eta_0 = \left(\eta_{01}, \eta_{02}\right)$,  $\eta_{01} \equiv \eta_{01}(X) = \mathbb{E}\left[\left. Y \right|X\right]$, $\eta_{02} \equiv \eta_{02}(X) = \mathbb{E}\left[\left. D \right|X\right]$. Let 
	$$
	m\left(W,\theta,\eta\right) = \mathbb{E}\left[\left.Y - \eta_1(X) - \theta\left(D - \eta_2(X)\right)\right|Z_1,X\right].
	$$

	We are interested in $\psi(\lambda_0) = \theta_0 \in \mathbb{R}$, whose Riesz representer is $r_\psi = \left(1,0\right)^{\prime} \in \overline{\Delta\left(\lambda_0\right)} \subseteq \mathbf{H} \equiv \mathbb{R} \times \mathcal{H}_\eta$, where $\mathcal{H}_\eta$ is a Hilbert space such that $ \mathcal{H}_\eta \subseteq L^2(X) \times L^2(X)$. In the restricted model, $\theta_0$ remains fixed and hence $h$ such that $\left<h,r_{\psi}\right>_{\mathbf{H}} = 0$ are those with $h = \left(0,b\right)$, i.e., $\delta = 0$. Therefore, it is suffices to focus on the derivative
	$$
	\nabla m \left[0,b\right] \equiv \mathbb{E}\left[\left.-b_1(X) + \theta_0 b_{2}(X)\right|Z_1,X\right] = -b_1(X) + \theta_0 b_2(X), \;\;\; b=\left(b_1,b_2\right).
	$$
	
	Our results from Section \ref{existencegral} imply that in this setting an OR-IV is a function in $L^2(Z_1,X)$ such that it is orthogonal to the range defined by the previous linear operator. The following proposition applies Theorem \ref{ThCMR} to this example. In particular, it gives a general expression for LR moments and gives necessary and sufficient conditions for their existence. 
	
	\begin{proposition}
		\label{theoremorthogonality}
		LR moments for model \eqref{eqmodel} can be constructed as follows 
		\begin{equation}
			\label{generallr}
			g\left(Z,\theta_0,\eta_0,\kappa_0\right) = \left(\tilde{Y} - \theta_0\tilde{D}\right)\kappa_0(Z_1,X), 
		\end{equation}
		where  $\kappa_0\left(Z_1,X\right) = \xi\left(Z_1,X\right) - \mathbb{E}\left[\left.\xi\left(Z_1,X\right)\right|X\right]$, with $\xi \in L^2(Z_1,X)$. Moreover, non-trivial LR moments exist iff $Z_1 \not\subset X$. 
	\end{proposition}
	
	Proposition \ref{theoremorthogonality} defines a general class of debiased moments for model \eqref{eqmodel}. In this context, an OR-IV can be seen as a recentered IV in the sense that its conditional expectation given pre-treatment characteristics is zero. Recentering IVs as a way to reduce bias is a strategy that has appeared in previous works, e.g., \cite{borusyak2020non}. Notice that we potentially have a large class of these, depending on the choice $\xi(Z_1,X)$. All of them lead to a debiased moment, suitable for proposing estimation procedures based on machine learning techniques below. As a particular example, when $\xi(Z_1,X) = Z_1$, we obtain the moment associated with DML of \cite{chernozhukov2018double}. Each of them defines a specific estimand, with particular identifying conditions. The next section characterizes a general class of relevant OR-IVs or ORR-IVs that lead to identification. 
	
	\subsection{Relevance}
	\label{srelevance}
	Our functional of interest is $\psi(\lambda_0) = \theta_0$, then, a relevant orthogonal moment is a moment satisfying 
	\begin{equation}
		\label{eqderiv}
		\frac{d}{d\tau} \mathbb{E}\left[\left(\tilde{Y} - \theta_\tau \tilde{D}\right) \kappa_0\left(Z_1,X\right)\right] = \delta \mathbb{E}\left[\left(D - \eta_{02}\left(X\right)\right)\kappa_0\left(Z_1,X\right)\right] \neq 0,
	\end{equation}
	where $\theta_\tau = \theta_0 + \tau \delta$. This is a relevance condition for our OR-IV $\kappa_0(Z_1,X)$. Note that in this example $\nabla m_{\psi}=\left(  \nabla m\right)  [r_\psi]=\mathbb{E}\left[\tilde{D}|Z\right]$, so (\ref{eqderiv}) is consistent with our general theorem on relevance, Theorem \ref{informative2}. This discussion immediately leads to the following result.
	\begin{proposition}
		\label{proprelevance}
		All relevant OR-IVs (ORR-IVs) for model \eqref{eqmodel} are based on a $\xi \in L^2(Z_1,X)$ such that 
		\begin{equation}
			\label{gralcondrel}
			\mathbb{E}\left[\left(D - \mathbb{E}\left[\left. D\right|X\right]\right)\left(\xi\left(Z_1,X\right) - \mathbb{E}\left[\left. \xi\left(Z_1,X\right)\right|X\right]\right)\right] \neq 0.
		\end{equation}
	\end{proposition}
	
	Next, we want to establish a mild sufficient condition such that if this holds we can obtain an ORR-IV whose $\xi$ satisfies \eqref{gralcondrel} and we demonstrate how to construct such $\xi$. 
	
	\begin{proposition}
		\label{relevancetheorem}
		Suppose that $\mathbb{E}\left[\left.D\right|Z_1,X\right] \neq  \mathbb{E}\left[\left.D\right|X\right]$.	 An ORR-IV in model \eqref{eqmodel}, $\kappa^*_0\left(Z_1,X\right)$, is such that 
		$$
		\kappa^*_0\left(Z_1,X\right) = \xi^{*}\left(Z_1,X\right) - \mathbb{E}\left[\left.\xi^{*}\left(Z_1,X)\right) \right|X\right],
		$$
		where 
		$$
		\xi^{*}\left(Z_1,X\right) = \left(\mathbb{E}\left[\left.D\right|Z_1,X\right] - \mathbb{E}\left[\left.D\right|X\right]\right) \nu\left(Z_1,X\right),
		$$
		for any $\nu \in L^2(Z_1,X)$ such that $\nu\left(Z_1,X\right) > 0$ a.s. 
	\end{proposition}
	
	Proposition \ref{relevancetheorem} implies that from any positive function $\nu$ a.s., we can construct an ORR-IV, provided that $\mathbb{E}\left[\left.D\right|Z_1,X\right] \neq  \mathbb{E}\left[\left.D\right|X\right]$, which is the relevance condition from the standard IV literature and can thus be deemed as mild. With all these choices, $\theta_0$ can be identified as
	
	\begin{equation}
		\label{theta}
		\theta_0  = \frac{\mathbb{E}\left[\tilde{Y}\kappa^{*}_0(Z_1,X)\right]}{\mathbb{E}\left[\tilde{D}\kappa^{*}_0(Z_1,X)\right]}  = \frac{\mathbb{E}\left[\tilde{Y}\xi^{*}\left(Z_1,X\right)\right]}{\mathbb{E}\left[\tilde{D}\xi^{*}\left(Z_1,X\right)\right]}.
	\end{equation}
	
	\bigskip Proposition \ref{relevancetheorem} says that there potentially exists a large class of OR-IVs that guarantees the relevance condition. However, there might be OR-IVs for which the relevance condition fails.  As an illustration, take $\nu(Z_1,X) = 1$ and suppose that $Z_1$ is a binary IV, say $Z_1 \in \left\{0,1\right\}$, then $\xi^{*}(Z_1,X) = 0$ is equivalent to $Cov\left(\left.D,Z_1\right|X\right) = \mathbb{E}\left[\left.\xi^{*}(Z_1,X)Z_1\right|X\right] = 0$ a.s. \citep[see Proposition 3.1 in][]{caetano2021identifying}. That means the minimal condition for identification of $\theta_0$ is that $D$ and $Z_1$ are correlated conditional on $X$. Next, suppose that we consider $\xi(Z_1,X) = Z_1$, as suggested by Double-Debiased Machine Learning estimator, so that $\kappa_0(Z_1,X) = Z_1 - \mathbb{E}\left[\left.Z_1 \right|X\right]$ and hence $\mathbb{E}\left[\tilde{D}\kappa_0(Z_1,X)\right] = \mathbb{E}\left[\tilde{D}Z_1\right] = \mathbb{E}\left[\xi^{*}(Z_1,X)Z_1\right]$. Thus, our choice $\xi^{*}(Z_1,X)$ can still lead to relevant orthogonal moments, while $Z_1$ does not, when $\mathbb{E}\left[\left.\xi^{*}(Z_1,X)Z_1\right|X\right] \neq 0$ but $\mathbb{E}\left[\xi^{*}(Z_1,X)Z_1\right]  = 0$. In our Monte Carlo experimentation below, we consider a data-generating process where such a situation occurs. Notice that with our choice $\xi^{*}$, the inner term of the expectation in the denominator of \eqref{theta} is always non-negative a.s., which is a necessary condition for obtaining an estimand with interpretable weights, as discussed by \cite{robustestimationcaetanoetal2023}. 
	
	Proposition \ref{relevancetheorem} potentially provides a large number of ORR-IVs, as they are indexed by a choice of a positive function $\nu$. If we were interested in efficiency the choice of $\nu$ matters, e.g., we might let $\nu(Z_1,X) = 1/\sigma(Z_1,X)^2$, where $\sigma(Z_1,X)^2 = \mathbb{E}\left[\left(\tilde{Y} - \theta_0\tilde{D}\right)^2\left.\right|Z_1,X\right]$, which is the reciprocal of the conditional variance of residuals given $Z_1$ and $X$; see \cite{chamberlain1992efficiency}. Since this is an unknown object, we will need to estimate it in practice, along with the other conditional expectations in $\xi^{*}$, which are also unknown.  
	
	To reduce the number of nuisance parameters, we simply let $\nu(Z_1,X)=1$ a.s. in our implementation below. Moreover, since 
	$$
	\mathbb{E}\left[\tilde{D} \kappa_0(Z_1,X)\right] = \mathbb{E}\left[\left(\mathbb{E}\left[\left.D \right|Z_1,X\right] - \mathbb{E}\left[\left.D \right|X\right]\right)\xi\left(Z_1,X\right)\right],
	$$
	for any OR-IV,  under the restriction that $\left(\mathbb{E}\left[\left.D \right|Z_1,X\right] - \mathbb{E}\left[\left.D \right|X\right]\right)\xi(Z_1,X) \geq 0$ a.s, the choice of $\xi$ that maximizes the correlation between $\left(\mathbb{E}\left[\left.D \right|Z_1,X\right] - \mathbb{E}\left[\left.D \right|X\right]\right)$ and $\xi$ (which can be seen as a measure of ``relevance") is  $\xi(Z_1,X) = c\xi^{*}(Z_1,X)$ with $\nu(Z_1,X)=1$ a.s., for some constant $c>0$; see Theorem 2.5 in \cite{robustestimationcaetanoetal2023}. Hence, our $\xi^{*}(Z_1,X)$ with $\nu(Z_1,X)=1$ leads to an ORR-IV that is proportional to the one that maximizes relevance.\footnote{In any IV setting, the scale of the IV does not affect identification: If $\kappa_0$ is the ``instrument" that yields a given estimand, then any other function that is proportional to $\kappa_0$ is also the ``instrument" of the same estimand.} Hereafter, we focus on an ORR-IV with this choice and when we discuss ``our estimand" below we refer to \eqref{theta} resulting from such a choice. More specifically, we focus on 
	\begin{equation}
		\label{theta1}
		\theta_0  = \frac{\mathbb{E}\left[\tilde{Y}\kappa^{*}_0(Z_1,X)\right]}{\mathbb{E}\left[\tilde{D}\kappa^{*}_0(Z_1,X)\right]}  = \frac{\mathbb{E}\left[\tilde{Y}\xi^{*}\left(Z_1,X\right)\right]}{\mathbb{E}\left[\tilde{D}\xi^{*}\left(Z_1,X\right)\right]}, \;\;\; \xi^{*}(Z_1,X) = \mathbb{E}\left[\left.D \right|Z_1,X\right] - \mathbb{E}\left[\left.D \right|X\right]. 
	\end{equation}
	\subsection{Does our estimand have a causal interpretation?}
	\label{sinterpretation}
	
	Assuming that the relevance condition in Proposition \ref{relevancetheorem} is satisfied, we now turn to an important point regarding the interpretation of \eqref{theta1}, under the standard assumptions of the LATE framework. 
	
	Remark that, in the previous development, we have not assumed anything regarding the support of the covariates of the model or how they relate (apart from Equation \eqref{eqmodel}). However, in this section, we restrict the variables $D$ and $Z_1$ to be binary variables. Then, we will show that \eqref{theta1} can be written as a convex combination of conditional LATEs, when model \eqref{eqmodel} is misspecified, i.e., its interpretation does not rely on $\mathbb{E}\left[\left. \varepsilon \right|Z_1,X\right] = 0$ being true. Hereafter, with some abuse of notation, let us denote the propensity score $p(Z_1,X) = \mathbb{E}\left[\left.D \right|Z_1,X\right]$, and $p_0 = p\left(Z_1=0,X\right)$ and $p_1 = p\left(Z_1=1,X\right)$.

	We revise the LATE framework of \cite{imbensangrist94}, \cite{angrist1995two}, and \cite{abadie2003semiparametric}. As usual, for every individual, let $Y(1)$ and $Y(0)$ be potential outcomes corresponding to the values of $Y$ that this individual would attain if treated ($D=1$) and if not treated $(D=0)$, respectively. The treatment effect is then $Y(1) - Y(0)$. Let $D(1)$ and $D(0)$ denote the potential treatment that correspond to the treatment that an individual receives when $Z_1=1$ and $Z_1=0$, respectively. Note $Y = Y(D)$ and $D = D(Z_1)$, i.e, only realized outcome and treatment status are observed. If the potential outcome were to depend on $Z_1$, we would write $Y = Y(D,Z_1)$.
	
	The population is composed of four sub-groups: Always-takers (with $D(1) = D(0) = 1$), Never-takers (with $D(1) = D(0) = 0$), Compliers (with $D(1) = 1$ and $D(0) =0$), and Defiers (with $D(1) = 0$ and $D(0) =0$). We follow \cite{kolesa2013estimation} and \cite{sloczynski2020should}, and define LATE as 
	$$
	\tau_{LATE} = \mathbb{E}\left[\left.Y(1) - Y(0)\right| D(1) \neq D(0)\right],
	$$
	which is the average treatment effect for individuals whose treatment status is affected by the instrument. Remark that the unconditional LATE can also be written as 
	$$
	\tau_{LATE} = \frac{\mathbb{E}\left[\pi(X) \tau(X)\right]}{\mathbb{E}\left[\pi(X)\right]},
	$$
	where 
	$$
	\tau(x) = \mathbb{E}\left[\left.Y(1) - Y(0)\right|D(1) \neq D(0), X = x\right],
	$$
	is the conditional LATE and 
	$$
	\pi(x) = \mathbb{P}\left(\left.D(1) \neq D(0) \right| X = x\right),
	$$
	is the conditional proportion of defiers and compliers. We next state the assumptions of the LATE framework. 
	\begin{assumption}
		\label{assumptioniv}\
		\begin{itemize}
			\item[(i)] (Conditional independence) $\left(Y(0,0), Y(0,1), Y(1,0), Y(1,1), D(0), D(1) \right) \perp Z_1 |X$; 
			\item[(ii)] (Exclusion restriction) $\mathbb{P}\left(\left.Y(1,d) = Y(0,d) \right| X\right) = 1$ for $d \in \{0,1 \}$ a.s.
			\item[(iii)] (Relevance) $0 < \mathbb{P}\left(\left.Z_1=1 \right| X\right) < 1$ and $\mathbb{P}\left(D(1) = 1|X \right) \neq \mathbb{P}\left(D(0) = 1|X \right)$ a.s.
		\end{itemize}
	\end{assumption}
	As is well-known, Assumption \ref{assumptioniv} is not enough to identify $\tau(X)$ and $\pi(X)$ (and thus $\tau_{LATE}$). It is also necessary to restrict the existence of defiers, as shown by \cite{imbensangrist94}. The following assumption was introduced by \cite{abadie2003semiparametric}. This rules out the existence of defiers at any value of covariates. 
	
	\begin{assumption}
		\label{SM}
		$\mathbb{P}\left(\left.D(1) \geq D(0)\right| X\right) = 1$ a.s.
	\end{assumption}
	Assumption \ref{SM} is referred to as a strong monotonicity assumption by \cite{sloczynski2020should}, as opposed to the weak monotonicity assumption:
	
	\begin{assumption}
		\label{WM}
		There exists a partition of the covariate space such that $\mathbb{P}\left(D(1) \geq D(0)|X\right) = 1$ a.s. in one subset and $\mathbb{P}\left(D(1) \leq D(0)|X\right) = 1$ a.s. on its complement. 
	\end{assumption}
	
	Assumption \ref{WM} is weaker than Assumption \ref{SM} since it does not rule out the existence of defiers at each value of the covariate space, but the coexistence of compliers and defiers at a given value of $X$. Thus, statements that are valid under Assumption \ref{WM} are automatically valid under Assumption \ref{SM}. As pointed out by \cite{sloczynski2020should}, considering heterogeneity in the way the instrument affects the treatment, depending on the values of covariates, is relevant in applied work.\footnote{For example, \cite{sloczynski2020should} considers the study by \cite{maestas2013does}, who analyze the effect of disability insurance benefit receipt on labor supply, exploiting exogenous variation in examiners' allowance rates as an instrument for benefit receipt. In the simplest case where there are two examiners, A and B, Assumption \ref{SM} would imply that A is always more sympathetic than B, regardless of applicants' characteristics. This is hard to believe. Instead, Assumption \ref{WM} would indicate that A is more lenient towards male candidates while B towards female candidates, for instance, which is a more plausible situation.}   
	
	Additionally, let 
	$$
	c(x) = sgn\left(\mathbb{P}\left( \left.D(1) \geq D(0) \right| X = x\right) - \mathbb{P}\left( \left.D(1) \leq D(0) \right| X = x\right)\right),
	$$
	where $sgn(\cdot)$ is the sign function. Note that $c(x)$ equals 1 if there are only compliers at $X=x$ and $-1$ if there are only defiers at $X = x$.  
	
	It is well-known that under Assumption \ref{assumptioniv} and \ref{WM} (and thus \ref{SM}), $\tau\left(X\right)$ is identified and equal to 
	$$
	\tau\left(x\right) = \frac{\mathbb{E}\left[\left. Y \right|Z_1=1,X=x\right] - \mathbb{E}\left[\left. Y \right|Z_1=0,X=x\right]}{\mathbb{E}\left[\left. D \right|Z_1=1,X=x\right] - \mathbb{E}\left[\left. D \right|Z_1=0,X=x\right]},
	$$
	\citep[see][]{imbensangrist94, sloczynski2020should}. Thus, $\tau(X)$ can be obtained as an IV estimand, constructed using $Z_1$ as an instrument in a population where $X$ is fixed. In addition, due to Assumption \ref{WM}, the instrument $Z_1$ enters the model only through the propensity score, and thus 
	$$
	\tau\left(x\right) = \frac{\mathbb{E}\left[\left. Y \right|p(x,Z_1) = p_1,X=x\right] - \mathbb{E}\left[\left. Y \right|p(x,Z_1) = p_0,X=x\right]}{\mathbb{E}\left[\left. D \right|Z_1=1,X=x\right] - \mathbb{E}\left[\left. D \right|Z_1=0,X=x\right]},
	$$
	\citep[see][]{kolesa2013estimation, heckman1999local, heckman2006understanding}. Furthermore, under the weak version of monotonicity, $\pi(X)$ can be identified as the absolute value of the denominator in $\tau(X)$.
	
	We now present the most important result of this section: $\theta_0$ in \eqref{theta1} can be written as a convex combination of $\tau(X)$ under the weaker version of monotonicity.
	
	\begin{proposition}
		\label{interpretationtheorem}
		Suppose that Assumptions \ref{assumptioniv} and \ref{WM} hold. Then, the estimand \eqref{theta1} can be written as 
		\begin{equation}
			\label{cmlw}
			\theta_{0} = \frac{\mathbb{E}\left[\omega(X) \tau(X)\right]}{\mathbb{E}\left[\omega(X)\right]}  \end{equation}
		where $\omega(x) = Var\left(\left.\mathbb{E}\left[D|Z_1,X\right]\right|X=x\right)$.
	\end{proposition}

	\bigskip The previous theorem shows that our estimand has a causal interpretation even if Assumption \ref{WM} is maintained, i.e., we do not need a stronger version of monotonicity to obtain a convex combination of conditional LATEs. Moreover, as shown by \cite{sloczynski2020should}, \eqref{cmlw} can also be expressed as 
	\begin{equation}
		\label{cmlw2}
		\theta_{0} = \frac{\mathbb{E}\left[\pi\left(X\right)^2 Var\left(\left. Z_1\right|X\right) \tau\left(X\right)\right]}{\mathbb{E}\left[\pi\left(X\right)^2 Var\left(\left. Z_1\right|X\right)\right]},
	\end{equation}
	and thus weights are always non-negative.
	
	The result in Theorem \ref{interpretationtheorem} is not new. The parameter $\theta_0$ coincides with the estimand of other estimators proposed by previous works, which have shown the causal interpretation of it. $\theta_0$ is the same estimand of the fully saturated two-stage least square estimator of \cite{angrist1995two} (Theorem 3). Unfortunately, the corresponding estimator proposed by \cite{angrist1995two} is obtained through a saturated model with discrete covariates and first-stage regressions of $D$ on interaction terms between the instrument and the covariates. This has limited the applicability of \cite{angrist1995two}'s estimator tremendously \citep[][]{sloczynski2020should}; see also the survey presented in \cite{blandhol2022tsls}. What is more, the moment that \cite{angrist1995two} considers is not LR. As we explain below, we propose estimating $\theta_0$ using machine learning tools, which do not involve running saturated models or discrete regressors, while standard inference is still valid as we use an orthogonal moment. \cite{kolesa2013estimation} proposes a (leave-one-out) unbiased jackknife instrumental variable estimator \citep[][]{phillips1977bias,angrist1999jackknife}. It has $\theta_0$ as its estimand and is consistent under many instrument asymptotics. The many instrument asymptotics is beyond the scope of this paper. Furthermore, the moment that \cite{kolesa2013estimation} considers coincides with the one in \cite{angrist1995two} and thus is not LR. 
	
	Moreover, $\xi^{*}(Z_1,X)$ is the optimal augmented instrument proposed by \cite{coussens2021improving}, $\left(p_1-p_0\right)\left(Z_1- \mathbb{E}\left[Z_1\right]\right)$, under the strong assumption that $Z_1 \perp \varepsilon, X$, with $E\left[\varepsilon X\right] = 0$, and under homoskedasticity. This instrument is optimal in the sense that it achieves the minimal asymptotic variance of an IV estimator.  In this case, our recommended ORR-IV captures the strength of the instrument, $\mathbb{E}\left[D|Z_1=1,X\right] - \mathbb{E}\left[D|Z_1=1,X\right] $, which is related to the probability of compliance with the instrument. \cite{coussens2021improving} argue that  their IV (and thus our preferred ORR-IV) can be interpreted as being weighing observations according to such probability. This idea mirrors the strategy discussed by \cite{abadie2024instrumental}, who propose an IV estimator that implicitly weights groups of observations according to the strength of the instrument in the first stage with a weighting scheme that has the same causal interpretation as \cite{kolesa2013estimation}'s (see Section 2.2.4 in \cite{abadie2024instrumental}). Similarly to us, \cite{coussens2021improving} recommends the use of machine learning to obtain the instrument. However, we do not assume the restricting independence between the instrument and covariates. Nevertheless, notice that even in situations where such an independence condition does not hold, the IV of \cite{coussens2021improving} is still valid and thus can be used to learn treatment effects.

	We finish this section by noticing, as Proposition \ref{interpretationtheorem} makes it clear, $\theta_0$ has an appealing interpretation, no matter if Assumption \ref{WM} or Assumption \ref{SM} holds. This is not true for all choices $\xi(Z_1,X)$. For example, consider $\xi\left(Z_1,X\right) = Z_1$ such that $\kappa_0(Z_1,X) = Z_1 - \mathbb{E}\left[\left.Z_1 \right|X\right] $, proposed by the DML. This is the standard IV estimator, applied in the context of machine learning. As such, under Assumptions \ref{assumptioniv} and \ref{WM},  it can be shown that its estimand can be written as 
	$$
	\theta_{DML} = \frac{\mathbb{E}\left[\left(\mathbb{E}\left[D|Z_1=1,X\right]  - \mathbb{E}\left[D|Z_1=0,X\right]\right) Var\left(Z_1|X\right) \tau(X)\right]}{\mathbb{E}\left[\left(\mathbb{E}\left[D|Z_1=1,X\right]  - \mathbb{E}\left[D|Z_1=0,X\right]\right)Var\left(Z_1|X\right)\right]}.
	$$
	Note that the weights for the case of $\theta_{DML}$ cannot be guaranteed to be non-negative under the weaker Assumption \ref{WM} as it depends on the term $\mathbb{E}\left[D|Z_1=1,X\right]  - \mathbb{E}\left[D|Z_1=0,X\right]$. Indeed, one can show that 
	$$
	\theta_{DML} = \frac{\mathbb{E}\left[c(X) \pi(X) Var\left(Z_1|X\right) \tau(X)\right]}{\mathbb{E}\left[c(X) \pi(X) Var\left(Z_1|X\right)\right]}, 
	$$
	\citep[see][Theorem 3.3]{sloczynski2020should}. As defiers are not ruled out, for some values of $X$, $c(X)$ might be negative. Therefore, a causal interpretation can only be attained under the stronger version of monotonicity when $c(X) = 1$ a.s. In contrast, the causal interpretation of our estimand $\theta_0$ remains valid even if defiers are not ruled out in the population, as long as Assumption \ref{WM} holds.

	\section{Estimation}
	\label{sestimation}
	
	As we indicated above, to estimate $\theta_0$ we use cross-fitting. That is, we randomly partition the sample intro $L$ groups, $I_{\ell}$, $\ell = 1,\cdots,L$, of equal size. Let  $\hat{\eta}_{1\ell}$ and $\hat{\eta}_{2\ell}$ be estimators of $\eta_{01}$ and $\eta_{02}$, respectively, using observations that are not in $I_\ell$. In addition, let $\hat{p}_{\ell}$ be an estimator of $p(Z_1,X) = \mathbb{E}\left[\left.D\right|Z_1,X\right]$, using observations that are not in $I_\ell$. Then, an estimator of $\theta_0$ is
	\begin{equation}
		\label{estimator}
		\hat{\theta}_{CML} = \frac{\frac{1}{n} \sum^{L}_{\ell = 1} \sum_{i \in I_\ell} \left(Y_{i} - \hat{\eta}_{1\ell}(X_i)\right)\hat{\kappa}_\ell\left(Z_i,X_i\right)}{\frac{1}{n} \sum^{L}_{\ell = 1} \sum_{i \in I_\ell} \left(D_i - \hat{\eta}_{2\ell}(X_i)\right)\hat{\kappa}_\ell\left(Z_i,X_i\right)},
	\end{equation}
	where $\hat{\kappa}_\ell\left(Z_{1i},X_i\right) = \hat{p}_\ell\left(Z_{1i},X_i\right) - \hat{\eta}_{2\ell}(X_i)$.\footnote{Note that $\mathbb{E}\left[\left.\xi^{*}\left(Z_1,X\right)\right|X\right] = 0$, and thus $\kappa_0\left(Z_1,X\right) = \xi^{*}\left(Z_1,X\right)$. Therefore, we will use these two expressions interchangeably in the sequel.} We called $\hat{\theta}_{CML}$ the \textit{Compliance Machine Learning Estimator (CML)} as it involves the estimation of the propensity score $p\left(Z_{1i},X_i\right)$, which equals the probability of complying with the treatment when $D$ is binary. We propose to estimate the conditional expectations in \eqref{estimator} with machine learning tools,  such as Lasso, random forest,  neural networks, boosting, and the like.  Note that, in a sense, $\hat{\theta}_{CML}$ can be seen as a simple IV estimator, where we consider a regression of $\left(Y - \hat{\eta}_1(X)\right)$ on $\left(D - \hat{\eta}_2(X)\right)$, using $\left(\hat{p} - \hat{\eta}_2(X)\right)$ as the instrument. As shown in Section \ref{sasymptotictheory}, inference on $\hat{\theta}_{CML}$ is standard. We finish this section with an algorithm to compute $\hat{\theta}_{CML}$: 
	
	\bigskip \noindent \textbf{Step 1:} Compute estimators $\hat{\eta}_{1\ell}$,  $\hat{\eta}_{2\ell}$, and $\hat{p}_\ell$, suing observations that are not in $I_\ell$, for each $\ell = 1,\cdots,L$.
	
	\bigskip \noindent \textbf{Step 2:} Generate $\hat{Y}_{i\ell} = Y_{i} - \hat{\eta}_{1\ell}(X_i)$, $\hat{D}_{i\ell} = D_i - \hat{\eta}_{2\ell}(X_i)$, $\hat{Z}_{1i\ell} = \hat{p}_\ell(Z_{1i},X_i) - \hat{\eta}_{2\ell}(X_i)$, $\ell = 1,\cdots,L$.
	
	\bigskip \noindent \textbf{Step 3:} Run an IV regression of $\hat{Y}_{i\ell}$ on $\hat{D}_{i\ell}$, using $\hat{Z}_{1i\ell}$ as an instrument, without intercept. Obtain $\hat{\theta}_{CML}$ as the coefficient of $\hat{D}_{i\ell}$. Asymptotically valid standard errors are shown in Section \ref{sasymptotictheory}. 
	
	\bigskip Remark that there exists an equivalent way to implement \eqref{estimator}.  $\hat{\theta}_{CML}$ can be implemented in the same fashion as the DML estimator, where we now have to use the estimated propensity score $\hat{p}$ instead of $Z_1$ as the instrument. However, the object $\mathbb{E}\left[\left.p\left(Z_1,X\right)\right|X\right]$ should be estimated through ``double cross-fitting". In particular, the observations used to estimate $p(Z_1,X)$ should be different than the one used to estimate $\mathbb{E}\left[\left.p\left(Z_1,X\right)\right|X\right]$, which in turn are not in $I_\ell$. In Sections \ref{smontecarlo} and \ref{soregon} we study if these two algorithms lead to different performances in fine samples.

	\section{Monte Carlo}
	\label{smontecarlo}
	This section presents a Monte Carlo experiment, which assesses the relative performance of CML against other competing alternatives, under different data generating processes (DGPs), and illustrates important points in our theory.
	
	We evaluate the behavior of CML in four DGPs. We consider the point that the conditions demanded by CML to identify $\theta_0$ are milder. In particular, they are more general than that of DML. As we indicated above,  when $Z \in \left\{0,1\right\}$, DML requires that $\mathbb{E}\left[Cov\left(\left.D,Z\right|X\right)\right] \neq 0$, which is not needed by CML as long as $\xi^{*}\left(Z,X\right) \neq 0$. Hence, in two of our DGPs, such an expectation is close to zero and the estimand of DML becomes large in absolute value. These will be DGP 1 and DGP 2. While CML can deal with these two situations, they are disadvantageous for DML. Hence, we consider two other DGPs such that the estimands of both coincide and are equal to zero. These are DGP 3 and DGP 4.

	To generate the previous DGPs we follow a similar strategy as in \cite{coussens2021improving}, Section 4. Let $\left(\delta_i, \varepsilon_i, \tau_i\right) \sim \mathcal{N}\left(0, \Sigma\right)$, where 
	\begin{equation}
		\label{Sigma}
		\Sigma = \begin{pmatrix}
			1 & \rho_{\delta \varepsilon} & \rho_{\delta \tau}\sigma_\tau \\
			\rho_{\delta\varepsilon} & 1 & \rho_{\tau \varepsilon}\sigma_\tau \\ 
			\rho_{\delta \tau}\sigma_\tau  & \rho_{\tau \varepsilon} \sigma_\tau & \sigma^2_\tau
		\end{pmatrix}.
	\end{equation}
	In this setting, $\delta_i$ is the latent tendency to receive treatment, $\tau_i$ is the treatment effect, and $\varepsilon_i$ is the baseline untreated potential outcome for individual $i$, as we show below. We also generate a regressor $X_{1}$ that can be either $X_{1i} = \bm{1}\left(\delta_i \geq 0\right)$ or $X_{1i} = \bm{1}\left(\delta_i > -\infty\right)$, depending on the DGP. Additionally, let $s_{1}$ and $s_{2}$ be numbers between 0 and 1 such that $1 - s_1 > s_2$. We obtain the potential treatment indicators as follows 
	\begin{equation}
		D_i(0) = \bm{1}\left(\Phi(\delta_i) > X_{1i}(1 - s_1) + (1-X_{1i}) s_2\right),\;\;\;\; D_i(1) = \bm{1}\left(\Phi(\delta_i) > X_{1i}s_2 + (1-X_{1i})(1-s_1)\right),
	\end{equation}
	where $s_1$ is the proportion of always-takers,  $s_2$ is the proportion of never-takers, and $\Phi$ is the standard normal cdf. Hence, $X_1$ determines if the strong monotonicity condition holds or not. Particularly, observe that when $X_{1i} = \bm{1}\left(\delta_i > - \infty \right)$, the strong monotonicity condition holds and defiers are ruled out in the entire population. In this case, $\mathbb{P}\left(\left.D_i(1) \geq D_i(0)\right|X\right) = 1$ a.s. However, when $X_{1i} = \bm{1}(\delta_i \geq 0)$, defiers are not ruled out in the population. In fact,  $\mathbb{P}\left(\left. D_i(0) \geq D_i(1) \right| X_{1i} = 0\right) = 1$. The outcome and treatment are 
	$$
	Y_{i} = D_i \tau_i + (1 + \alpha \delta_i)\varepsilon_i,\;\;\;\;\; D_i = D_i(0)(1-Z_{1i}) + D_i(1)Z_{1i},
	$$
	where $\alpha$ determines the degree of heterogeneity present in the variance of the untreated outcome, and $Z_{1i} \in \left\{0, 1\right\}$. Note that when $\rho_{\delta \varepsilon} \neq 0$  in \eqref{Sigma}, treatment $D$ is endogenous, and the use of IVs is needed. Additionally, if $\rho_{\delta \tau} \neq 0$, treatment is heterogeneous in the population, i.e., $\tau(X)$ is an actual function of $X$. We generate a second predictor, $X_{2i} \sim \mathcal{N}\left(0, \sigma^2_X\right)$, and let $X = \left(X_{1}, X_{2}\right)$. Notice that the inclusion of this second variable in the vector $X$ will only add noise in the estimation of the conditional expectations, as it does not have any predictive power on the rest of the variables of the model.  
	
	In our context, when  $X_{1i} = \bm{1}(\delta_i \geq 0)$, for $D_{i}(1) \neq D_{i}(0)$, it can be shown that
	\begin{equation}
		\label{eidentificationcond}
		\mathbb{E}\left[Cov\left(D_i,Z_{1i}|X_i\right) \right]  = \mathbb{P}\left(X_{1i} = 1\right)Var\left(Z_{1i}|X_{1i} = 1\right)  - \mathbb{P}\left(X_{1i} = 0\right)Var\left(Z_{1i}|X_{1i} =0\right).
	\end{equation}
	Since $\mathbb{P}\left(X_{1} = 1\right) = \mathbb{P}\left(X_{1} = 0\right)$, the above implies that $\mathbb{E}\left[Cov\left(D,Z_1|X\right) \right] = 0$ if the conditional variance of $Z_1$ is constant. We consider cases where the previous expectation is close to zero, making the denominator of the estimand of CML close to zero. Particularly, we let $\mathbb{P}\left(Z_1=1 |X_1\right) = \Phi\left(\alpha_Z + \beta_{XZ}X_1\right)$. Moreover, we have $\beta_{XZ} = 0.5$. Additionally, we let $\beta_{XZ} = 0.001$, and thus the identifiability of $\theta_{DML}$ becomes more difficult. As we show in the Appendix, when $X_{1i} = \bm{1}\left(\delta_i > -\infty\right)$, $\tau\left(X_i\right) = 0$, and thus $\tau_{LATE} = \theta_0 = \theta_{DML} = 0$. In this case, even though \eqref{eidentificationcond} is close to zero, $\theta_{DML}$ is well defined and coincides with the probability limit of CML. This is a more fair situation for DML.  We summarize our DGPs in the following table: 
	
	\begin{table}[H]
		\centering
		\caption{DGPs considered in Monte Carlo}
		\begin{tabular}{l|c|cc|l}
			\cline{2-4}
			& \multirow{2}{*}{$X_1$} & \multicolumn{2}{c|}{$Var(Z_1|X_1)$} &  \\ \cline{3-4}
			&  & \multicolumn{1}{c|}{$\beta_{XZ} = 0.5$} & $\beta_{XZ} = 0.001$ &  \\ \cline{2-4}
			& $X_1 = \bm{1}\left(\delta_i \geq 0\right)$ & \multicolumn{1}{c|}{DGP1} & DGP2 &  \\ \cline{2-4}
			& $X_1 = \bm{1}\left(\delta_i > -\infty\right)$ & \multicolumn{1}{c|}{DGP3} & DGP4 &  \\ \cline{2-4}
		\end{tabular}
	\end{table}
	
	In all DGPs, $\sigma_\tau = 1$, $\rho_{\delta \tau} = 0.5$, $\rho_{\delta \varepsilon} =0.5$, $\rho_{\tau \varepsilon} = 1$, $\alpha = 0$, $\sigma_X = 2$, $\alpha_z = 0$, $s_1 = 0.2$, and $s_2=0.4$. Our experimentation is always based on 1,000 Monte Carlo repetitions.

	We compute CML in two different versions. First, as suggested by the algorithm of Section \ref{sestimation}. Second, we implement CML similarly to DML, but with $\hat{p}\left(Z_{1i},X_i\right)$ as the IV, using double cross-fitting (DC) to estimate the conditional expectation of $\hat{p}$ given $X$. To illustrate the severity of not accounting for first-stage bias, i.e., not using a LR moment, we compute the estimator that would result from the moment proposed by \cite{angrist1995two} (AI). This is a moment similar to  \eqref{generallr}, evaluated at $\xi^{*}(Z_{1i},X_i)$, but using $Y_{i}$ and $D_i$, instead of $\tilde{Y}_{1i}$ and $\tilde{D_i}$, respectively \citep[cf. Equation (8) in][]{angrist1995two}. Finally, we report results with DML. In each of the previous cases, conditional expectations are estimated using two machine learning tools, namely, random forest (RF) and neural networks (NN).\footnote{All simulations were conducted in R. For RF, we use the function \texttt{ranger}; for NN, we use the function \texttt{neuralnet}.} Cross-fitting is always implemented, with $L = 4$. 
	
	Tables \ref{table:MSE_500}-\ref{table:MSE_2000} report the performance of the previous estimators, under each of the four DGPs. MSEs are displayed, along with their decomposition in terms of squared bias and variance. As target parameters, we consider the corresponding estimands (under the LATE framework) and the LATE estimand $\tau_{LATE}$. Note that the estimator based on the moment suggested by \cite{angrist1995two} has as its probability limit \eqref{cmlw2}, the same estimand of CML. The estimand of DML is $\theta_{DML}$. What is more,  in DGP 3 and DGP 4, all the estimands coincide and are equal to $\tau_{LATE}$.  
	
	We can draw several conclusions from our Monte Carlo exercise. First, not employing an orthogonal moment has serious consequences in terms of the performance of the resulting estimator. Under DGP 1 and DGP 2, AI reports larger biases across the Monte Carlo experiments than CML. Our simulations also suggest a gain in terms of variance when debiased moments are used. This is explained by the fact that a non-orthogonal estimator is more sensitive to first-stage estimations than one that is based on a debiased moment, as such, variability in the first-stage estimation would be translated into variability of the estimator, provoking that the resulting estimations are ``less continuous" with respect to the data. That the variance of a non-LR estimator is higher than CML also results in a poorer performance when the LATE estimand is the target parameter. As the second halves of the tables report, AI presents larger MSEs than CML. This is interesting since even though the estimand of CML is different from the LATE parameter, it can still be closer to it, in comparison with naive estimators that do not account for first-stage biases. The main driver of this is the significantly larger variance of a non-LR estimator. A last advantage of using LR moments is related to the fact that the particular machine learning tool used in estimation does not seem to play an important role, for a sufficiently large sample size. If we compare the performance of CML using RF or NN from Table \ref{table:MSE_2000}, we observe that the differences between them, in terms of bias and variance, are negligible.  However, if we compare columns 1 and 5 of Table \ref{table:MSE_2000}, for DGP 1 and DGP 2, we observe that the performance of AI crucially depends on whether RF or NN is employed, this choice matters not only in terms of bias but also in terms of variance. In particular, we see that when neural networks are used for estimating the first-stage, AI presents a performance significantly superior than when random forest is employed.

	Second, in this setting, the bias coming from the first stage will not necessarily be sufficiently large to affect inference. The panels associated with DGP 3 and DGP 4 show small bias and variance for AI, and indeed they are similar to the ones reported by CML. The reason behind this is that in those DGPs, covariates in $X$ are not relevant whatsoever, and they are completely independent of the rest of variables in the model. Since the nuisance parameters are functions of $X$, a non-LR moment is orthogonal already. This is a very unrealistic setting, but helps us to show that it is always preferable to work with an orthogonal moment than one that is not, as bias and variance can be kept under control, uniformly in the DGP.

	Third, having derived a LR moment based on an IV that satisfies the relevance condition in general circumstances is convenient. Under DGP 1, when DML is close to failing its identifiability requirement and the estimand is different from zero, the performance of DML is seriously compromised and is worse than CML. DML tends to report a larger bias than CML. This deteriorates even more when the identifiability of CML is more ``fragile", with DGP 2. In addition, the variance of DML is systematically higher. As we indicated above, these situations are very counterproductive for DML, which will have associated an almost unbounded estimand. However, when we consider $\tau_{LATE}$ as the target parameter for both estimators, we observe that still DML has higher MSEs than CML. Once more, DML reports larger bias than CML. Note that what also plays a part in these results is the fact that in those DGPs the strong monotonicity condition fails. Therefore, while $\tau_{LATE}$ and \eqref{cmlw2} have positive weights, $\theta_{DML}$ presents negative weights. This makes the probability limit of CML considerably depart from the true LATE effect (see Section \ref{sadditionalmc}).  In contrast, in DGP 3 and DGP 4,  both estimators have the same asymptotic limit. In these situations, CML and DML present similar performances, regarding bias. Nevertheless, as indicated by Table \ref{table:MSE_500}, CML has to pay a higher price in terms of variance. This is expected as CML deals with more nuisance parameters than DML. However, as the sample sizes increase, the variance reported by CML tends to be of the same order as that of DML. The main takeaway of this discussion is that CML is an appealing estimator, as its relevance condition is likely to be satisfied in more general contexts than other LR estimators based on other choice of IV. This is translated into a better-performance estimator, in terms of bias and variance, in different DGPs.     
	
	Lastly, we assess whether the particular algorithm to implement DML is relevant. To this end, we compare the performances of CML and CML (DC). Table \ref{table:MSE_500} shows that while DML (DC) reports smaller bias, it presents larger variance. This is not surprising, as CML (DC) deals with higher uncertainty in estimation. Nevertheless, as the sample size increases, both estimators behave almost identically, as suggested by Table \ref{table:MSE_1000} and \ref{table:MSE_2000}. In particular, Table \ref{table:MSE_2000} reports almost identical MSEs for both estimators, as they are similar in terms of bias and variance. Overall, we conclude that the way CML is implemented does not seem to influence the performance of the estimator, for a sufficiently large sample size, at least for these DGPs.

	\begin{table}[H]
		\scriptsize
		\addtolength{\tabcolsep}{-1pt}
		\centering 
		\caption{MSE decomposition ($n = 500$)}
		\label{table:MSE_500}
		\begin{threeparttable}
			\begin{tabular}{cllllllll}
				\hline \hline
				& \multicolumn{8}{l}{$\;\;\;\;\;\;\;\;\;\;\;\;\;\;\;\;\;\;\;\;\;\;\;\;\;\;\;\;\;\;\;\;\;\;\;\;\;\;\;\;\;\;\;\;\;\;\;\;\;\;\;\;\;\;\;\;\;\;\;\;\;\;$\textsc{Own Estimand}}\\
				& AI & DML & CML & CML (DC) & AI & DML & CML & CML (DC) \\ & (RF) & (RF) & (RF) & (RF) & (NN) & (NN) & (NN) & (NN) \\ \hline
				& \multicolumn{8}{l}{$\;\;\;\;\;\;\;\;\;\;\;\;\;\;\;\;\;\;\;\;\;\;\;\;\;\;\;\;\;\;\;\;\;\;\;\;\;\;\;\;\;\;\;\;\;\;\;\;\;\;\;\;\;\;\;\;\;\;\;\;\;\;\;\;\;\;\;\;\;\;\;\;$\textsc{DGP 1}} \\ \hline
				MSE & 15.077 $\times 10^4$  & 44.851 & 0.178 & 0.208 & 2.548 & 44.852 & 0.358 & 0.541 \\ 
				Bias$^2$ & 2.479 & 44.106 & 0.026 & 0.026 & 0.015 & 44.133 & 0.026 & 0.015 \\ 
				Var & 15.077 $\times 10^4$  & 0.746 & 0.152 & 0.182 & 2.534 & 0.719 & 0.333 & 0.526 \\ \hline
				& \multicolumn{8}{l}{$\;\;\;\;\;\;\;\;\;\;\;\;\;\;\;\;\;\;\;\;\;\;\;\;\;\;\;\;\;\;\;\;\;\;\;\;\;\;\;\;\;\;\;\;\;\;\;\;\;\;\;\;\;\;\;\;\;\;\;\;\;\;\;\;\;\;\;\;\;\;\;\;$\textsc{DGP 2}} \\ \hline
				MSE & 76.358 $\times 10^3$ & 24.674 $\times 10^{11}$ & 0.144 & 0.169 & 1.856 & 24.674$\times 10^{11}$ & 0.287 & 0.608 \\ 
				Bias$^2$ & 91.940 & 24.674 $\times 10^{11}$  & 0.020 & 0.014 & 0.014 & 24.674$\times 10^{11}$ & 0.026 & 0.016 \\ 
				Var & 76.266 $\times 10^3$  & 0.442 & 0.123 & 0.155 & 1.842 & 0.439 & 0.261 & 0.592 \\ \hline
				& \multicolumn{8}{l}{$\;\;\;\;\;\;\;\;\;\;\;\;\;\;\;\;\;\;\;\;\;\;\;\;\;\;\;\;\;\;\;\;\;\;\;\;\;\;\;\;\;\;\;\;\;\;\;\;\;\;\;\;\;\;\;\;\;\;\;\;\;\;\;\;\;\;\;\;\;\;\;\;$\textsc{DGP 3}} \\ \hline
				MSE & 0.137 & 0.135 & 0.138 & 0.205 & 0.143 & 0.137 & 0.145 & 0.160 \\ 
				Bias$^2$ & 0.012 & 0.016 & 0.014 & 0.001 & 0.012 & 0.016 & 0.013 & 0.003 \\ 
				Var & 0.125 & 0.119 & 0.124 & 0.205 & 0.131 & 0.120 & 0.132 & 0.157 \\ \hline
				& \multicolumn{8}{l}{$\;\;\;\;\;\;\;\;\;\;\;\;\;\;\;\;\;\;\;\;\;\;\;\;\;\;\;\;\;\;\;\;\;\;\;\;\;\;\;\;\;\;\;\;\;\;\;\;\;\;\;\;\;\;\;\;\;\;\;\;\;\;\;\;\;\;\;\;\;\;\;\;$\textsc{DGP 4}} \\ \hline
				MSE & 0.123 & 0.120 & 0.124 & 0.155 & 0.127 & 0.123 & 0.127 & 0.136 \\ 
				Bias$^2$ & 0.012 & 0.016 & 0.014 & 0.000 & 0.012 & 0.016 & 0.012 & 0.004 \\ 
				Var & 0.111 & 0.104 & 0.110 & 0.155 & 0.115 & 0.107 & 0.115 & 0.132 \\   
				& \multicolumn{8}{l}{$\;\;\;\;\;\;\;\;\;\;\;\;\;\;\;\;\;\;\;\;\;\;\;\;\;\;\;\;\;\;\;\;\;\;\;\;\;\;\;\;\;\;\;\;\;\;\;\;\;\;\;\;\;\;\;\;\;\;\;\;\;\;$\textsc{LATE Estimand}}\\
				& AI & DML & CML & CML (DC) & AI & DML & CML & CML (DC) \\ & (RF) & (RF) & (RF) & (RF) & (NN) & (NN) & (NN) & (NN) \\ \hline
				& \multicolumn{8}{l}{$\;\;\;\;\;\;\;\;\;\;\;\;\;\;\;\;\;\;\;\;\;\;\;\;\;\;\;\;\;\;\;\;\;\;\;\;\;\;\;\;\;\;\;\;\;\;\;\;\;\;\;\;\;\;\;\;\;\;\;\;\;\;\;\;\;\;\;\;\;\;\;\;$\textsc{DGP 1}} \\ \hline
				MSE & 15.0767 $\times 10^4$ & 0.849 & 0.169 & 0.199 & 2.542 & 0.824 & 0.349 & 0.534 \\ 
				Bias$^2$ & 2.580 & 0.103 & 0.017 & 0.017 & 0.008 & 0.105 & 0.017 & 0.008 \\ 
				Var & 15.0767 $\times 10^4$ & 0.746 & 0.152 & 0.182 & 2.534 & 0.719 & 0.333 & 0.526 \\ \hline
				& \multicolumn{8}{l}{$\;\;\;\;\;\;\;\;\;\;\;\;\;\;\;\;\;\;\;\;\;\;\;\;\;\;\;\;\;\;\;\;\;\;\;\;\;\;\;\;\;\;\;\;\;\;\;\;\;\;\;\;\;\;\;\;\;\;\;\;\;\;\;\;\;\;\;\;\;\;\;\;$\textsc{DGP 2}} \\ \hline
				MSE & 76.358 $\times 10^3$ & 0.533 & 0.144 & 0.169 & 1.856 & 0.530 & 0.287 & 0.608 \\ 
				Bias$^2$ & 91.940 & 0.090 & 0.020 & 0.014 & 0.014 & 0.091 & 0.026 & 0.016 \\ 
				Var & 76.266 $\times 10^3$ & 0.442 & 0.123 & 0.155 & 1.842 & 0.439 & 0.261 & 0.592 \\ \hline
				& \multicolumn{8}{l}{$\;\;\;\;\;\;\;\;\;\;\;\;\;\;\;\;\;\;\;\;\;\;\;\;\;\;\;\;\;\;\;\;\;\;\;\;\;\;\;\;\;\;\;\;\;\;\;\;\;\;\;\;\;\;\;\;\;\;\;\;\;\;\;\;\;\;\;\;\;\;\;\;$\textsc{DGP 3}} \\ \hline
				MSE & 0.137 & 0.135 & 0.138 & 0.205 & 0.143 & 0.137 & 0.145 & 0.160 \\ 
				Bias$^2$ & 0.012 & 0.016 & 0.014 & 0.001 & 0.012 & 0.016 & 0.013 & 0.003 \\ 
				Var & 0.125 & 0.119 & 0.124 & 0.205 & 0.131 & 0.120 & 0.132 & 0.157 \\  \hline
				& \multicolumn{8}{l}{$\;\;\;\;\;\;\;\;\;\;\;\;\;\;\;\;\;\;\;\;\;\;\;\;\;\;\;\;\;\;\;\;\;\;\;\;\;\;\;\;\;\;\;\;\;\;\;\;\;\;\;\;\;\;\;\;\;\;\;\;\;\;\;\;\;\;\;\;\;\;\;\;$\textsc{DGP 4}} \\ \hline
				MSE & 0.123 & 0.120 & 0.124 & 0.155 & 0.127 & 0.123 & 0.127 & 0.136 \\ 
				Bias$^2$ & 0.012 & 0.016 & 0.014 & 0.000 & 0.012 & 0.016 & 0.012 & 0.004 \\ 
				Var & 0.111 & 0.104 & 0.110 & 0.155 & 0.115 & 0.107 & 0.115 & 0.132 \\   
				\hline \hline
			\end{tabular}
			\begin{tablenotes}
				\scriptsize
				\item NOTE: The table shows the MSE decomposition of different estimators, considering as target parameters the corresponding estimand and the LATE estimand $\tau_{LATE}$. AI is an estimator based on the moment recommended by \cite{angrist1995two} (Equation 8); CML is the Double-Debiased-Machine Learning Estimator of \cite{chernozhukov2018double}; CML is the Compliance Machine Learning Estimator; CML (DC) is the Compliance Machine Learning Estimator that has been implemented similarly to DML, using double cross-fitting (DC). Two different machine learning tools were used to estimate all the conditional expectations: (1) random forest (RF), and (2) neural networks (NN). Results are based on $1,000$ Monte Carlo repetitions. 
			\end{tablenotes}   
		\end{threeparttable}
	\end{table}

	\begin{table}[H]
		\scriptsize
		\addtolength{\tabcolsep}{-1pt}
		\centering 
		\caption{MSE decomposition ($n = 1,000$)}
		\label{table:MSE_1000}
		\begin{threeparttable}
			\begin{tabular}{cllllllll}
				\hline \hline
				& \multicolumn{8}{l}{$\;\;\;\;\;\;\;\;\;\;\;\;\;\;\;\;\;\;\;\;\;\;\;\;\;\;\;\;\;\;\;\;\;\;\;\;\;\;\;\;\;\;\;\;\;\;\;\;\;\;\;\;\;\;\;\;\;\;\;\;\;\;$\textsc{Own Estimand}}\\
				& AI & DML & CML & CML (DC) & AI & DML & CML & CML (DC) \\ & (RF) & (RF) & (RF) & (RF) & (NN) & (NN) & (NN) & (NN) \\ \hline
				& \multicolumn{8}{l}{$\;\;\;\;\;\;\;\;\;\;\;\;\;\;\;\;\;\;\;\;\;\;\;\;\;\;\;\;\;\;\;\;\;\;\;\;\;\;\;\;\;\;\;\;\;\;\;\;\;\;\;\;\;\;\;\;\;\;\;\;\;\;\;\;\;\;\;\;\;\;\;\;$\textsc{DGP 1}} \\ \hline
				MSE & 24.739 $\times 10^3$ & 44.786 & 0.102 & 0.110 & 0.515 & 44.774 & 0.155 & 0.162 \\ 
				Bias$^2$ & 0.530 & 44.470 & 0.035 & 0.037 & 0.037 & 44.459 & 0.028 & 0.023 \\ 
				Var &  24.739 $\times 10^3$ & 0.316 & 0.067 & 0.074 & 0.478 & 0.315 & 0.127 & 0.139 \\  \hline
				& \multicolumn{8}{l}{$\;\;\;\;\;\;\;\;\;\;\;\;\;\;\;\;\;\;\;\;\;\;\;\;\;\;\;\;\;\;\;\;\;\;\;\;\;\;\;\;\;\;\;\;\;\;\;\;\;\;\;\;\;\;\;\;\;\;\;\;\;\;\;\;\;\;\;\;\;\;\;\;$\textsc{DGP 2}} \\ \hline
				MSE & 32.377 $\times 10^3$ & 24.674 $\times 10^{11}$ & 0.083 & 0.089 & 2.216 & 24.674 $\times 10^{11}$& 0.133 & 0.143 \\ 
				Bias$^2$ & 178.897 & 24.674 $\times 10^{11}$ & 0.024 & 0.021 & 0.008 & 24.674 $\times 10^{11}$ & 0.021 & 0.016 \\ 
				Var & 32197.978 & 0.213 & 0.059 & 0.068 & 2.208 & 0.212 & 0.112 & 0.128 \\  \hline
				& \multicolumn{8}{l}{$\;\;\;\;\;\;\;\;\;\;\;\;\;\;\;\;\;\;\;\;\;\;\;\;\;\;\;\;\;\;\;\;\;\;\;\;\;\;\;\;\;\;\;\;\;\;\;\;\;\;\;\;\;\;\;\;\;\;\;\;\;\;\;\;\;\;\;\;\;\;\;\;$\textsc{DGP 3}} \\ \hline
				MSE & 0.075 & 0.075 & 0.075 & 0.080 & 0.076 & 0.076 & 0.076 & 0.076 \\ 
				Bias$^2$ & 0.011 & 0.013 & 0.012 & 0.002 & 0.012 & 0.013 & 0.012 & 0.006 \\ 
				Var & 0.064 & 0.062 & 0.063 & 0.078 & 0.064 & 0.063 & 0.064 & 0.069 \\ \hline
				& \multicolumn{8}{l}{$\;\;\;\;\;\;\;\;\;\;\;\;\;\;\;\;\;\;\;\;\;\;\;\;\;\;\;\;\;\;\;\;\;\;\;\;\;\;\;\;\;\;\;\;\;\;\;\;\;\;\;\;\;\;\;\;\;\;\;\;\;\;\;\;\;\;\;\;\;\;\;\;$\textsc{DGP 4}} \\ \hline
				MSE & 0.073 & 0.072 & 0.073 & 0.073 & 0.072 & 0.072 & 0.073 & 0.071 \\ 
				Bias$^2$ & 0.013 & 0.015 & 0.014 & 0.004 & 0.013 & 0.015 & 0.014 & 0.009 \\ 
				Var & 0.059 & 0.057 & 0.059 & 0.069 & 0.059 & 0.057 & 0.059 & 0.063 \\   
				& \multicolumn{8}{l}{$\;\;\;\;\;\;\;\;\;\;\;\;\;\;\;\;\;\;\;\;\;\;\;\;\;\;\;\;\;\;\;\;\;\;\;\;\;\;\;\;\;\;\;\;\;\;\;\;\;\;\;\;\;\;\;\;\;\;\;\;\;\;$\textsc{LATE Estimand}}\\
				& AI & DML & CML & CML (DC) & AI & DML & CML & CML (DC) \\ & (RF) & (RF) & (RF) & (RF) & (NN) & (NN) & (NN) & (NN) \\ \hline
				& \multicolumn{8}{l}{$\;\;\;\;\;\;\;\;\;\;\;\;\;\;\;\;\;\;\;\;\;\;\;\;\;\;\;\;\;\;\;\;\;\;\;\;\;\;\;\;\;\;\;\;\;\;\;\;\;\;\;\;\;\;\;\;\;\;\;\;\;\;\;\;\;\;\;\;\;\;\;\;$\textsc{DGP 1}} \\ \hline
				MSE & 24.739 $\times 10^3$ & 0.437 & 0.091 & 0.099 & 0.504 & 0.436 & 0.145 & 0.154 \\ 
				Bias$^2$ & 0.577 & 0.122 & 0.025 & 0.025 & 0.026 & 0.121 & 0.018 & 0.015 \\ 
				Var & 24.739 $\times 10^3$ & 0.316 & 0.067 & 0.074 & 0.478 & 0.315 & 0.127 & 0.139 \\  \hline
				& \multicolumn{8}{l}{$\;\;\;\;\;\;\;\;\;\;\;\;\;\;\;\;\;\;\;\;\;\;\;\;\;\;\;\;\;\;\;\;\;\;\;\;\;\;\;\;\;\;\;\;\;\;\;\;\;\;\;\;\;\;\;\;\;\;\;\;\;\;\;\;\;\;\;\;\;\;\;\;$\textsc{DGP 2}} \\ \hline
				MSE & 32.377 $\times 10^3$ & 0.308 & 0.083 & 0.089 & 2.216 & 0.307 & 0.133 & 0.143 \\ 
				Bias$^2$ & 178.897 & 0.095 & 0.024 & 0.021 & 0.008 & 0.095 & 0.021 & 0.016 \\ 
				Var & 32.198 $\times 10^3$ & 0.213 & 0.059 & 0.068 & 2.208 & 0.212 & 0.112 & 0.128 \\  \hline
				& \multicolumn{8}{l}{$\;\;\;\;\;\;\;\;\;\;\;\;\;\;\;\;\;\;\;\;\;\;\;\;\;\;\;\;\;\;\;\;\;\;\;\;\;\;\;\;\;\;\;\;\;\;\;\;\;\;\;\;\;\;\;\;\;\;\;\;\;\;\;\;\;\;\;\;\;\;\;\;$\textsc{DGP 3}} \\ \hline
				MSE & 0.075 & 0.075 & 0.075 & 0.080 & 0.076 & 0.076 & 0.076 & 0.076 \\ 
				Bias$^2$ & 0.011 & 0.013 & 0.012 & 0.002 & 0.012 & 0.013 & 0.012 & 0.006 \\ 
				Var & 0.064 & 0.062 & 0.063 & 0.078 & 0.064 & 0.063 & 0.064 & 0.069 \\   \hline
				& \multicolumn{8}{l}{$\;\;\;\;\;\;\;\;\;\;\;\;\;\;\;\;\;\;\;\;\;\;\;\;\;\;\;\;\;\;\;\;\;\;\;\;\;\;\;\;\;\;\;\;\;\;\;\;\;\;\;\;\;\;\;\;\;\;\;\;\;\;\;\;\;\;\;\;\;\;\;\;$\textsc{DGP 4}} \\ \hline
				MSE & 0.073 & 0.072 & 0.073 & 0.073 & 0.072 & 0.072 & 0.073 & 0.071 \\ 
				Bias$^2$ & 0.013 & 0.015 & 0.014 & 0.004 & 0.013 & 0.015 & 0.014 & 0.009 \\ 
				Var & 0.059 & 0.057 & 0.059 & 0.069 & 0.059 & 0.057 & 0.059 & 0.063 \\  
				\hline \hline
			\end{tabular}
			\begin{tablenotes}
				\scriptsize
				\item NOTE: The table shows the MSE decomposition of different estimators, considering as target parameters the corresponding estimand and the LATE estimand $\tau_{LATE}$. AI is an estimator based on the moment recommended by \cite{angrist1995two} (Equation 8); CML is the Double-Debiased-Machine Learning Estimator of \cite{chernozhukov2018double}; CML is the Compliance Machine Learning Estimator; CML (DC) is the Compliance Machine Learning Estimator that has been implemented similarly to DML, using double cross-fitting (DC). Two different machine learning tools were used to estimate all the conditional expectations: (1) random forest (RF), and (2) neural networks (NN). Results are based on $1,000$ Monte Carlo repetitions. 
			\end{tablenotes}   
		\end{threeparttable}
	\end{table}

	\begin{table}[H]
		\scriptsize
		\addtolength{\tabcolsep}{-1pt}
		\centering 
		\caption{MSE decomposition ($n = 2,000$)}
		\label{table:MSE_2000}
		\begin{threeparttable}
			\begin{tabular}{cllllllll}
				\hline \hline
				& \multicolumn{8}{l}{$\;\;\;\;\;\;\;\;\;\;\;\;\;\;\;\;\;\;\;\;\;\;\;\;\;\;\;\;\;\;\;\;\;\;\;\;\;\;\;\;\;\;\;\;\;\;\;\;\;\;\;\;\;\;\;\;\;\;\;\;\;\;$\textsc{Own Estimand}}\\
				& AI & DML & CML & CML (DC) & AI & DML & CML & CML (DC) \\ & (RF) & (RF) & (RF) & (RF) & (NN) & (NN) & (NN) & (NN) \\ \hline
				& \multicolumn{8}{l}{$\;\;\;\;\;\;\;\;\;\;\;\;\;\;\;\;\;\;\;\;\;\;\;\;\;\;\;\;\;\;\;\;\;\;\;\;\;\;\;\;\;\;\;\;\;\;\;\;\;\;\;\;\;\;\;\;\;\;\;\;\;\;\;\;\;\;\;\;\;\;\;\;$\textsc{DGP 1}} \\ \hline
				MSE & 23.486 $\times 10^5$ & 44.652 & 0.071 & 0.074 & 36.285 & 44.666 & 0.099 & 0.101 \\ 
				Bias$^2$ & 21.336 $\times 10^2$ & 44.514 & 0.040 & 0.041 & 0.196 & 44.529 & 0.034 & 0.031 \\ 
				Var & 23.465 $\times 10^5$ & 0.138 & 0.032 & 0.033 & 36.089 & 0.137 & 0.065 & 0.070 \\ \hline
				& \multicolumn{8}{l}{$\;\;\;\;\;\;\;\;\;\;\;\;\;\;\;\;\;\;\;\;\;\;\;\;\;\;\;\;\;\;\;\;\;\;\;\;\;\;\;\;\;\;\;\;\;\;\;\;\;\;\;\;\;\;\;\;\;\;\;\;\;\;\;\;\;\;\;\;\;\;\;\;$\textsc{DGP 2}} \\ \hline
				MSE & 26.475 $\times 10^3$ & 24.674 $\times 10^{11}$ & 0.058 & 0.059 & 0.273 & 24.674 $\times 10^{11}$  & 0.095 & 0.098 \\ 
				Bias$^2$ & 19.815 & 24.674 $\times 10^{11}$ & 0.031 & 0.029 & 0.024 & 24.674 $\times 10^{11}$  & 0.026 & 0.023 \\ 
				Var & 26.475 $\times 10^3$ & 0.100 & 0.028 & 0.029 & 0.249 & 0.099 & 0.069 & 0.075 \\  \hline
				& \multicolumn{8}{l}{$\;\;\;\;\;\;\;\;\;\;\;\;\;\;\;\;\;\;\;\;\;\;\;\;\;\;\;\;\;\;\;\;\;\;\;\;\;\;\;\;\;\;\;\;\;\;\;\;\;\;\;\;\;\;\;\;\;\;\;\;\;\;\;\;\;\;\;\;\;\;\;\;$\textsc{DGP 3}} \\ \hline
				MSE & 0.042 & 0.043 & 0.043 & 0.040 & 0.042 & 0.043 & 0.043 & 0.041 \\ 
				Bias$^2$ & 0.014 & 0.015 & 0.015 & 0.009 & 0.014 & 0.015 & 0.014 & 0.011 \\ 
				Var & 0.028 & 0.028 & 0.028 & 0.031 & 0.028 & 0.028 & 0.028 & 0.029 \\  \hline
				& \multicolumn{8}{l}{$\;\;\;\;\;\;\;\;\;\;\;\;\;\;\;\;\;\;\;\;\;\;\;\;\;\;\;\;\;\;\;\;\;\;\;\;\;\;\;\;\;\;\;\;\;\;\;\;\;\;\;\;\;\;\;\;\;\;\;\;\;\;\;\;\;\;\;\;\;\;\;\;$\textsc{DGP 4}} \\ \hline
				MSE & 0.042 & 0.042 & 0.042 & 0.039 & 0.042 & 0.042 & 0.042 & 0.040 \\ 
				Bias$^2$ & 0.016 & 0.017 & 0.017 & 0.011 & 0.016 & 0.017 & 0.016 & 0.013 \\ 
				Var & 0.026 & 0.025 & 0.026 & 0.028 & 0.026 & 0.026 & 0.026 & 0.027 \\ 
				& \multicolumn{8}{l}{$\;\;\;\;\;\;\;\;\;\;\;\;\;\;\;\;\;\;\;\;\;\;\;\;\;\;\;\;\;\;\;\;\;\;\;\;\;\;\;\;\;\;\;\;\;\;\;\;\;\;\;\;\;\;\;\;\;\;\;\;\;\;$\textsc{LATE Estimand}}\\
				& AI & DML & CML & CML (DC) & AI & DML & CML & CML (DC) \\ & (RF) & (RF) & (RF) & (RF) & (NN) & (NN) & (NN) & (NN) \\ \hline
				& \multicolumn{8}{l}{$\;\;\;\;\;\;\;\;\;\;\;\;\;\;\;\;\;\;\;\;\;\;\;\;\;\;\;\;\;\;\;\;\;\;\;\;\;\;\;\;\;\;\;\;\;\;\;\;\;\;\;\;\;\;\;\;\;\;\;\;\;\;\;\;\;\;\;\;\;\;\;\;$\textsc{DGP 1}} \\ \hline
				MSE & 23.486 $\times 10^5$ & 0.262 & 0.060 & 0.062 & 36.258 & 0.262 & 0.088 & 0.090 \\ 
				Bias$^2$ & 21.307$\times 10^2$ & 0.124 & 0.028 & 0.030 & 0.169 & 0.125 & 0.023 & 0.021 \\ 
				Var & 23.465 $\times 10^5$ & 0.138 & 0.032 & 0.033 & 36.089 & 0.137 & 0.065 & 0.070 \\   \hline
				& \multicolumn{8}{l}{$\;\;\;\;\;\;\;\;\;\;\;\;\;\;\;\;\;\;\;\;\;\;\;\;\;\;\;\;\;\;\;\;\;\;\;\;\;\;\;\;\;\;\;\;\;\;\;\;\;\;\;\;\;\;\;\;\;\;\;\;\;\;\;\;\;\;\;\;\;\;\;\;$\textsc{DGP 2}} \\ \hline
				MSE & 26.474 $\times 10^3$ & 0.207 & 0.058 & 0.059 & 0.273 & 0.207 & 0.095 & 0.098 \\ 
				Bias$^2$ & 19.815 & 0.107 & 0.031 & 0.029 & 0.024 & 0.108 & 0.026 & 0.023 \\ 
				Var & 26.455 $\times 10^3$ & 0.100 & 0.028 & 0.029 & 0.249 & 0.099 & 0.069 & 0.075 \\   \hline
				& \multicolumn{8}{l}{$\;\;\;\;\;\;\;\;\;\;\;\;\;\;\;\;\;\;\;\;\;\;\;\;\;\;\;\;\;\;\;\;\;\;\;\;\;\;\;\;\;\;\;\;\;\;\;\;\;\;\;\;\;\;\;\;\;\;\;\;\;\;\;\;\;\;\;\;\;\;\;\;$\textsc{DGP 3}} \\ \hline
				MSE & 0.042 & 0.043 & 0.043 & 0.040 & 0.042 & 0.043 & 0.043 & 0.041 \\ 
				Bias$^2$ & 0.014 & 0.015 & 0.015 & 0.009 & 0.014 & 0.015 & 0.014 & 0.011 \\ 
				Var & 0.028 & 0.028 & 0.028 & 0.031 & 0.028 & 0.028 & 0.028 & 0.029 \\   \hline
				& \multicolumn{8}{l}{$\;\;\;\;\;\;\;\;\;\;\;\;\;\;\;\;\;\;\;\;\;\;\;\;\;\;\;\;\;\;\;\;\;\;\;\;\;\;\;\;\;\;\;\;\;\;\;\;\;\;\;\;\;\;\;\;\;\;\;\;\;\;\;\;\;\;\;\;\;\;\;\;$\textsc{DGP 4}} \\ \hline
				MSE & 0.042 & 0.042 & 0.042 & 0.039 & 0.042 & 0.042 & 0.042 & 0.040 \\ 
				Bias$^2$ & 0.016 & 0.017 & 0.017 & 0.011 & 0.016 & 0.017 & 0.016 & 0.013 \\ 
				Var & 0.026 & 0.025 & 0.026 & 0.028 & 0.026 & 0.026 & 0.026 & 0.027 \\ 
				\hline \hline
			\end{tabular}
			\begin{tablenotes}
				\scriptsize
				\item NOTE: The table shows the MSE decomposition of different estimators, considering as target parameters the corresponding estimand and the LATE estimand $\tau_{LATE}$. AI is an estimator based on the moment recommended by \cite{angrist1995two} (Equation 8); CML is the Double-Debiased-Machine Learning Estimator of \cite{chernozhukov2018double}; CML is the Compliance Machine Learning Estimator; CML (DC) is the Compliance Machine Learning Estimator that has been implemented similarly to DML, using double cross-fitting (DC). Two different machine learning tools were used to estimate all the conditional expectations: (1) random forest (RF), and (2) neural networks (NN). Results are based on $1,000$ Monte Carlo repetitions. 
			\end{tablenotes}   
		\end{threeparttable}
	\end{table}

	\section{The Oregon Health Insurance Experiment}
	\label{soregon}
	
	In 2008,  Oregon expanded its coverage for Medicaid (the U.S.
	social program that provides health insurance to disadvantaged people who
	cannot afford private insurance). The state conducted lottery drawings to
	randomly select names from a waiting list of almost 90,000 uninsured adults,
	as demand far exceeded supply. Selected participants were given the
	opportunity (for themselves and any household member) to apply for Medicaid
	and, conditional on having their application approved, enroll in the program;
	for a detailed description of the experiment see \cite{finkelstein2012oregon}.
	
	Not all participants who were selected through the lottery were ultimately
	enrolled in Medicaid.\footnote{According to \cite{finkelstein2012oregon}, about 30\% of the individuals selected by the lottery were successfully enrolled.} This raises concerns about the exogeneity of receiving health insurance as treatment. Those who were enrolled might have characteristics that might be correlated with target outcomes, which would contaminate the true effect of the treatment, even after controlling for important covariates. Nevertheless, the fact that potential candidates were randomly drawn from the lottery makes it available an arguably valid instrument that can be employed to uncover the causal effect of being enrolled in Medicaid. This is what we are exploiting in this analysis.   
	
	We obtained the data from a survey that was conducted approximately one year after insurance coverage began to all individuals selected by the lottery and a roughly equal number of non-selected individuals in seven different waves. The data contains 23,741 subjects.\footnote{Observe that this is the total number of individuals who participated in the survey.} This is the same data used in \cite{finkelstein2012oregon} for their survey results. In the experiment, randomization was conducted at the household level. Instrument $Z_{1i}$ is an indicator variable that is equal to one if the household where individual $i$ lives was selected in the lottery draw. $D_i$ is the main measure of treatment in \cite{finkelstein2012oregon}, an indicator variable for whether the individual was enrolled in Medicaid.\footnote{This variable is called ``Ever on Medicaid" in \cite{finkelstein2012oregon}.}  We consider two sets of observable characteristics in this study. $X_1$ is a vector of variables that is always considered and includes indicator variables for household size.\footnote{Larger households were significantly over-represented among those that were selected by the lottery; see footnote 12 in \cite{finkelstein2012oregon}.} It also includes indicator variables for survey wave as the proportion of treated individuals is not constant across waves. In addition, we include interaction terms between household size and survey wave indicators. Then, we expand this set of covariates with $X_2$, which includes sex, age, whether English is the language preferred by the subject to conduct the survey, race, education, employment status, and income. In each of our results reported below, standard errors were clustered at the household level.\footnote{We remark that the standard errors that we report are the one obtained by the ``off-the-shelf" routines, which might not coincide with the one recommended by our asymptotic theory (cf. Section \ref{sasymptotictheory} in the Appendix). We report them only for comparison reasons.} Moreover, we weighted observations according to the survey weights provided to account for the sampling design of the survey.    
	
	We are interested in learning the treatment effect of health insurance coverage on health care utilization. This is an important aspect since an increased health care utilization would be translated into larger health costs from the supply side.\footnote{Additional use of health care services might bring improved health status as well; however, a one-year span might not large enough to detect these positive effects.} We study this through eight different outcomes. We divide them into two groups. On one side, we analyze extensive margins or access to health care services. On the other side, we focus on the intensity of the use of such services. These are the same outcomes considered in Table V of \cite{finkelstein2012oregon}. 
	
	\cite{finkelstein2012oregon} work with two models: a simple OLS model of the outcomes on $Z_1$ (and $X_1$), in which case, intention-to-treat (ITT) results are reported, and an IV model where now $D$ is instrumented by $Z_1$ (we denote this LATE model, following the original terminology in \cite{finkelstein2012oregon}). In contrast, we consider a flexible model on covariates $X$, as the one reported in \eqref{eqmodel}, where nuisance parameters are estimated using machine learning tools. In this study, we use random forest (RF) and neural nerworks (NN). Apart from DML, we report the estimation of one additional estimator that uses the IV suggested by \cite{coussens2021improving} (CS), namely, $\xi\left(Z_1,X\right) = \left(\mathbb{E}\left[\left. D \right|Z_1=1,X\right] - \mathbb{E}\left[\left. D \right|Z_1=1,X\right]\right) \left(Z_1 - \mathbb{E}\left[Z_1\right]\right)$. As in our Monte Carlo exercise,  we implement our CML estimator in two ways. Additionally, we augment the set of covariates considered by \cite{finkelstein2012oregon} by including $X_2$. 
	
	The estimation results are displayed in Table \ref{table:HC}. The columns that do not include the additional controls (add. controls) are the original results in \cite{finkelstein2012oregon}. They report non-negligible and  statistically significant increases in prescription drugs and outpatient visits in health centers. For instance, considering the LATE estimates, their results would indicate that insurance is associated with a 0.35 (std. err. = 0.18) increase in the number of prescription drugs currently taken, which is equivalent to a 15\% increase, and a 1.08 (std err. = 0.18) in the number of outpatient visits, or a 55\% increase. They do not find statistically significant effects on emergency room (ER) visits, on either margin. Furthermore,  their findings are not supportive of any impact on inpatient hospital admissions, on either margin. These findings remain virtually the same when additional covariates $X_2$ are included in the analysis, although now we observe a significant 0.04 (std err. = 0.024) increase on the extensive margin of ER visits, which represents a 15\% increase. 
	
	We now turn to the estimates based on machine learning tools. Relative to the previous baseline estimates, the four machine learning estimators, using the two machine learning tools, report qualitatively the same effect on drug prescriptions and outpatient visits, i.e., health insurance access has a positive effect on these. Nevertheless, while DML mostly reports smaller point estimates, CS and CML suggest larger effects. For instance, they would indicate that the treatment increases the number of drug prescriptions by roughly 0.4 (std err. = 0.17), which is 0.05 units more than that in \cite{finkelstein2012oregon}. Similarly, they would suggest that health insurance coverage increases outpatient visits by 1.3 (std err. = 0.19/0.196 (CS) and 0.172/0.190 (CML)), which represents a 0.22 larger effect than that in \cite{finkelstein2012oregon}. These effects decrease to 0.255/0.278 (std err.=0.175/0.174) and 1.075/1.133 (std err.=0.190/0.195), respectively, when DML is considered. This reduction might respond to the failure of the strong monotonicity assumption in this setting, which leads to the estimand of DML to have negative weights. Additionally, machine learning supports the finding that a positive impact on ER visits is observed, on either margin, although the effects are more statistically significant for the extensive margin, and these effects are in line with the simple IV model with additional covariates. The use of a given machine learning algorithm does not seem to influence our findings as the estimates based on RF and NN are always similar. What is more, we point out that CML, in most cases, reports more precise estimates than any other machine learning estimator. Although, the gains in terms of efficiency are not large in this case, possibly due to the considerable number of observations present in the study. Finally, the data on this application indicates that there do not seem to be qualitatively and quantitatively relevant differences between the two implementations of CML. The only remarkable difference appears for current drugs prescribed as CML (DC) reports that access to Medicaid increases the number of them by 0.63/0.507 (std err.=0.208/0.152), which are 0.23/0.12 units larger than that suggested by CML.

	\begin{sidewaystable}
		\begin{table}[H]
			\scriptsize
			\addtolength{\tabcolsep}{-1pt}
			\centering 
			\caption{Health Care Utilization (Survey Data)}
			\label{table:HC}
			\begin{threeparttable}
				\begin{tabular}{clllllllllllll}
					\hline \hline
					& \multicolumn{13}{l}{$\;\;\;\;\;\;\;\;\;\;\;\;\;\;\;\;\;\;\;\;\;\;\;\;\;\;\;\;\;\;\;\;\;\;\;\;\;\;\;\;\;\;\;\;\;\;\;\;\;\;\;\;\;\;\;\;\;\;\;\;\;\;\;\;\;\;\;\;\;\;\;\;\;\;\;\;\;\;\;\;\;\;\;\;\;\;\;$\textsc{Extensive margin (any)}}\\
					& Control & ITT & LATE & ITT & LATE & DML & CS & CML & CML (DC) & DML & CS & CML & CML (DC) \\ & mean &  &  &  &  & (RF) & (RF) & (RF) & (RF) &  (NN) & (NN) & (NN) & (NN) \\ \hline   
					
					Prescription drugs  & 0.637 & 0.025*** & 0.088*** & 0.023*** & 0.08*** & 0.081*** & 0.079*** & 0.087*** & 0.07** & 0.08*** & 0.084*** & 0.081*** & 0.077*** \\ 
					(currently) & (0.481) & (0.008) & (0.029) & (0.008) & (0.029) & (0.029) & (0.027) & (0.026) & (0.033) & (0.029) & (0.027)& (0.026) & (0.024) \\ 
					Outpatient visits & 0.574 & 0.062*** & 0.212*** & 0.058*** & 0.2*** & 0.19*** & 0.236*** & 0.222*** & 0.236*** & 0.2*** & 0.235*** & 0.234*** & 0.23*** \\ 
					(lst. 6 mths.) & (0.494) & (0.007) & (0.025) & (0.008) & (0.026) & (0.027) & (0.025) & (0.024) & (0.029) & (0.026) & (0.025) & (0.024) & (0.022) \\ 
					ER visits & 0.261 & 0.006  & 0.022  & 0.012** & 0.04** & 0.041** & 0.038* & 0.039** & 0.045* & 0.043** & 0.041** & 0.045** & 0.035** \\ 
					(lst. 6 mths.)  6 & (0.439) & (0.007) & (0.023) & (0.007) & (0.024) &  (0.024) & (0.024) & (0.023) & (0.028) & (0.024) & (0.024) & (0.024) & (0.021) \\ 
					Inpatient hospital & 0.072 & 0.002  & 0.008 & 0.003 & 0.009 & 0.006  & 0.021* & 0.014 & 0.018 & 0.011  & 0.014 & 0.017  & 0.019* \\ 
					admissions (lst. 6 mths.) & (0.259) & (0.004) & (0.014)  & (0.004) & (0.014) & (0.014) & (0.015) & (0.014) & (0.016) & (0.014) & (0.015) & (0.014) & (0.012) \\ 
					
					& \multicolumn{12}{l}{$\;\;\;\;\;\;\;\;\;\;\;\;\;\;\;\;\;\;\;\;\;\;\;\;\;\;\;\;\;\;\;\;\;\;\;\;\;\;\;\;\;\;\;\;\;\;\;\;\;\;\;\;\;\;\;\;\;\;\;\;\;\;\;\;\;\;\;\;\;\;\;\;\;\;\;\;\;\;\;\;\;\;\;\;\;\;\;$\textsc{Total utilization (number)}}\\
					& Control & ITT & LATE & ITT & LATE & DML & CS & CML & CML (DC) & DML & CS & CML & CML (DC) \\ & mean &  &  &  &  & (RF) & (RF) & (RF) & (RF) &  (NN) & (NN) & (NN) & (NN) \\ \hline   
					
					Prescription drugs  & 2.318 & 0.1** & 0.347** & 0.079* & 0.275* & 0.255* & 0.397** & 0.409*** & 0.628*** & 0.278* & 0.393** & 0.382** & 0.507*** \\ 
					(currently)   2 & (2.878) & (0.051) & (0.176) & (0.051) & (0.174) & (0.175) & (0.171) & (0.16) & (0.208) & (0.174) & (0.172) & (0.17) & (0.152) \\ 
					Outpatient visits & 1.914 & 0.314*** & 1.083*** & 0.324*** & 1.11*** & 1.075*** & 1.294*** & 1.319*** & 1.342*** & 1.133*** & 1.273*** & 1.261*** & 1.133*** \\ 
					(lst. 6 mths.) & (3.087) & (0.054) & (0.182) & (0.058) & (0.195) & (0.19) & (0.19) & (0.172) & (0.183) & (0.195) & (0.196) & (0.19) & (0.151) \\ 
					ER visits & 0.47 & 0.007  & 0.026  & 0.02  & 0.07 & 0.081* & 0.083* & 0.089* & 0.125** & 0.08* & 0.071  & 0.082* & 0.074* \\ 
					(lst. 6 mths.) & (1.037) & (0.016) & (0.056) & (0.017) & (0.058) & (0.058)& (0.06) & (0.056) & (0.064) & (0.058) & (0.061) & (0.06) & (0.05) \\ 
					Inpatient hospital & 0.097 & 0.006 & 0.021 & 0.007  & 0.024  & 0.016  & 0.031* & 0.019  & 0.03  & 0.02  & 0.024  & 0.024  & 0.023 \\ 
					admissions (lst. 6 mths.) & (0.4) & (0.006) & (0.021) & (0.006)  & (0.021) & (0.022) & (0.024) & (0.022)  & (0.026)  & (0.022)  & (0.024) & (0.022)  & (0.02)\\ 
					&  &  &  &  &  &  &  &  &  &  &  \\ 
					Add. controls &  & No & No & Yes & Yes &  Yes & Yes & Yes & Yes & Yes & Yes & Yes & Yes \\ 
					\hline \hline
				\end{tabular}
				\begin{tablenotes}
					\scriptsize
					\item NOTE: Standard errors are reported in parenthesis. Hospital admissions exclude childbirth. ITT refers to intention to treat, where the estimation comes from an OLS model of the outcome on $Z_1$, an indicator variable for being selected in the lottery draw. The rest of the columns is an IV estimation, where $D$ is "ever on Medicaid" during the period of analysis, which is the same treatment used by \cite{finkelstein2012oregon}. LATE reports estimates of a simple IV procedure. DML is the Double Debiased Machine Learning Estimator of \cite{chernozhukov2018double}. CS uses as instrument the one suggested by \cite{coussens2021improving}. DML is our estimator, and DML (DC) is our estimator that has been implemented similarly to DML, using double cross-fitting (DC). These last four were estimated using random forest (RF) and Neural Networks (NN). All models include household size effects, survey wave fixed effects, and the interaction between the two. Additional controls (Add. controls) include sex, age, whether English is the language preferred by the subject on the survey, race, education, employment status, and income. All standard errors are clustered at the household level. Observations are always weighted by survey weights. Our data is composed of 23,741 survey respondents.
				\end{tablenotes}   
			\end{threeparttable}
		\end{table}
	\end{sidewaystable}

	\section{Final Remarks}
	\label{sconclusion}
	
	This paper has developed a characterization of debiased moments for regular models defined by general semiparametric CMRs with possibly different conditioning variables. Moreover, it has provided necessary and sufficient conditions for such characterization, and studied the situations where our construction can lead to moments that can detect departures of the parameter of interest from the truth at a parametric rate in a local sense. Such results apply to a broad class of smooth functionals of finite and infinite dimensional parameters. Then, our results are potentially useful in extending the application of LR moments to complex models with CMRs, which are ubiquitous in economics and statistics. 
	
	As an application of our theory, this work has derived an appealing estimator for treatment effects in the relevant setting of a partially linear model with endogeneity. This estimator is based on a particular choice of OR-IVs, the ORR-IV. This function measures the strength of excluded exogenous characteristics on the treatment, given pre-treatment variables. This choice is convenient as it leads to an estimator that enjoys three important properties in the standard LATE framework: (1) local robustness to first-stage estimation, (2) a general relevance condition, and (3) a meaningful causal interpretation. We propose the use of machine learning algorithms to compute such an IV, and the resulting CML estimator. Our numerical experimentation shows the satisfactory performance of such estimator in finite samples, across different DGPs. We have also considered real-world data to illustrate the applicability of CML. We believe that (1)-(3) are sufficiently relevant reasons to consider CML as an arguably appealing IV treatment effect estimator. We hope that it will provide applied researchers with an additional tool to uncover important causal effects, in general contexts.

	The above is a particular application of our theory, where it was relatively simple to estimate OR-IVs. In other contexts, it might not be the case. Therefore, it would be interesting to propose general algorithms that allow researchers to estimate those OR-IVs in a wide range of settings, where it might be challenging to obtain a closed-form expression for those functions. This will lead to the computation of debiased moments in a data-driven or automatic way. Such a procedure would extend the scope of application of the proposed characterization to other contexts, relevant for applied work. We leave this point for future research.

	\newpage
	\begin{center}
		{\Huge{ \textsc{Appendix}}}
	\end{center}
	
	\begin{appendix}
		
		\section{Proofs of Main Results}
		\label{proofs}
		
		\noindent\textbf{Proof of Theorem \ref{ThCMR}:} 
		Define
		\[
		M_{\psi_{0}^{\perp}}:=\left(  \nabla m\right)  \circ\Pi_{r_{\psi}^{\perp}%
		}\qquad\text{and}\qquad M_{\psi_{0}^{\perp}}^{\ast}:=\Pi_{r_{\psi}^{\perp}%
		}\circ\left(  \nabla m\right)  ^{\ast},
		\]
		where $\left(  \nabla m\right)  ^{\ast}$ is the adjoint operator of $\nabla
		m$ and $\Pi_{r^{\perp}_\psi}$ denotes the orthogonal projection onto the orthocomplement of the linear span of $\left(r_{\psi,t}\right)^{d_\psi}_{t=1}$. By Theorem 1 in \cite{arganaraz2023existence}, a LR moment must be an element of the orthogonal complement of the tangent space generated by scores of mappings $\tau \mapsto \lambda_\tau$ within the restricted model that has $\psi\left(\lambda_\tau\right) = \psi_0$ fixed. Theorem 4.1 in \cite{chen2018overidentification} yields that the orthocomplement of the tangent space of the full model
		is
		\[
		\overline{T}^{\perp}=\{g\in L_{0}^{2}:g(Z,\lambda_{0}, \kappa_0)=%
		{\textstyle\sum\nolimits_{j=1}^{J}}
		\rho_{j}(Z,\lambda_{0})\kappa_{0j}(W_{j})\text{ for }\kappa_{0}=(\kappa
		_{0j})_{j=1}^{J}\in\overline{\mathcal{R}(\nabla m)}^{\perp}\}.
		\]
		We are not concerned with the full model in this paper, but rather the
		restricted model where deviations satisfy $ \dot{\psi}(h) = \langle r_{\psi},h\rangle
		_{\mathbf{H}}=0.$ For these deviations $h=\Pi_{r_{\psi}^{\perp}}h.$ Thus, the
		result follows by restricting the domain of $\nabla m$ to directions for which
		$h=\Pi_{r_{\psi}^{\perp}}h.$ Next, use duality arguments \citep[see][p. 157]{luenberger1997optimization}
		to conclude that $\kappa_0 = 0$, i.e., the only possible LR moment is $g = 0$, is equivalent to $\overline{\mathcal{V}}_{r^{\perp}_\psi} =\overline{\mathcal{R}(M_{\psi
				_{0}^{\perp}})}^{\perp}=\mathcal{N}(M_{\psi_{0}^{\perp}}^{\ast})=\{0\}.$
		Furthermore, that $\kappa_0 \in \mathcal{N}\left(M^{*}_{\psi^{\perp}_0}\right)$ is equivalent to $\Pi_{r_{\psi}^{\perp}}\circ(\nabla m)^{\ast}\kappa_0=0$, which  is
		equivalent to $(\nabla m)^{\ast}\kappa_0=\sum_{j=1}^{d_{\psi}}C_{j}r_{\psi,j}$
		for constants $C_{j}.$ $\blacksquare$\bigskip
		
		\noindent\textbf{Proof of Theorem \ref{informative2}:} We decompose $h=\Pi_{r_{\psi}}h+\Pi_{r_{\psi}^{\perp}}h$. Then, note that by the law of iterated expectations and orthogonality, 
		\begin{align}
			\frac{d}{d\tau} \mathbb{E}\left[g\left(Z,\theta_\tau,\eta_\tau,\kappa_0\right)\right] & = \mathbb{E}\left[\left(\nabla m\right)^{\prime}[h]\kappa_0(Z)\right] \nonumber \\ & = \mathbb{E}\left[\left(\nabla m\right)^{\prime}[\Pi_{r_{\psi}}h]\kappa_0(Z)\right], \label{mm2}			
		\end{align}
		where notice that  $\left(  \nabla m\right)
		[\Pi_{r_{\psi}}h]$ is in the linear span of $\nabla m_{\psi}=\left(  \nabla
		m\right)  [r_{\psi}].$ Thus, \eqref{mm2} is equal to zero (yielding no information for the parameter of interest) if
		$\nabla m_{\psi}$ is orthogonal to $\kappa_0=(\kappa_{0j})_{j=1}^{J}%
		\in\mathcal{N}(M_{\psi_{0}^{\perp}}^{\ast}).$
		$\blacksquare$\bigskip

		\bigskip \noindent \textbf{Proof of Proposition \ref{theoremorthogonality}:} The conclusion of the proposition follows by Theorem \ref{ThCMR}. In particular, observe that 
		
		\begin{align}
			\frac{d}{d\tau} \mathbb{E}\left[g\left(Z,\theta_0,\eta_0 + \tau b,\kappa_0\right)\right] & = \frac{d}{d \tau} \mathbb{E}\left[\left(Y - \eta_{01}(X) - \tau b_1(X) - \theta_0\left(D - - \eta_{02}(X) - \tau b_2(X)\right)\right)\kappa_0(Z_1,X)\right] \nonumber \\ & = \mathbb{E}\left[\frac{d}{d\tau} \mathbb{E}\left[\left. Y - \eta_{01}(X) - \tau b_1(X) - \theta_0\left(D - \eta_{02}(X) - \tau b_2(X)\right)\right|Z_1,X\right]\kappa_0\left(Z_1,X\right)\right] \nonumber \\ & = \mathbb{E}\left[\left.\mathbb{E}\left[-b_1(X) + \theta_0b_2(X)\right|Z_1,X\right]\kappa_0(Z_1,X)\right] \nonumber \\ & = \mathbb{E}\left[\left(-b_1(X) + \theta_0b_2(X)\right)\kappa_0(Z_1,X)\right] \nonumber \\ & = 0.
			\label{orthogonalitycond}
		\end{align}
		where the first equality follows by construction, the second equality holds by the law of iterated expectations, the third equality comes from the chain rule, the fourth equality is direct, and the last one is due to orthogonality. Hence, \eqref{cond1} holds. Furthermore, using the law of iterated expectation, it is easy to see that  \eqref{cond2} also holds. Then, we observe that a necessary and sufficient condition for the existence of non-trivial LR moments is that  $Z_1 \not\subset X$. $\blacksquare$
		
		\bigskip \noindent \textbf{Proof of Proposition \ref{proprelevance}:} The results follows directly from expression \eqref{eqderiv} which characterizes the relevance of LR moments, constructed in Proposition \ref{theoremorthogonality},  within model \eqref{eqmodel}. $\blacksquare$

		\bigskip \noindent \textbf{Proof of Proposition \ref{relevancetheorem}:} By orthogonality and the law of iterated expectations, 
		\begin{align}
			\mathbb{E}\left[\left(D - \mathbb{E}\left[\left.D\right|,X\right]\right)\kappa^{*}_0\left(Z_1,X\right)\right] & = \mathbb{E}\left[\left(D - \mathbb{E}\left[\left.D\right|,X\right]\right)\xi^{*}\left(Z_1,X\right)\right] \nonumber \\ & = \mathbb{E}\left[\left(\mathbb{E}\left[\left.D \right|Z_1,X\right] - \mathbb{E}\left[\left.D \right|X\right] \right)^2\nu\left(Z_1,X\right)\right] \label{ortcondresult}.
		\end{align}
		Since $\left(\mathbb{E}\left[\left.D \right|Z_1,X\right] - \mathbb{E}\left[\left.D \right|X\right] \right)^2\nu\left(Z_1,X\right) >0$ if $\mathbb{E}\left[\left.D\right|Z_1,X\right] \neq  \mathbb{E}\left[\left.D\right|X\right]$, we conclude that \eqref{ortcondresult} is necessarily different from zero.  $\blacksquare$

		\bigskip \noindent \textbf{Proof of Proposition \ref{interpretationtheorem}:} As $\mathbb{E}\left[\left.p(Z_1,X) \right|X\right] = \mathbb{E}\left[\left. D\right|X\right]$, by orthogonality, 
		$$
		\theta_{0} = \frac{\mathbb{E}\left[Y \xi^{*}\left(Z_1,X\right)\right]}{\mathbb{E}\left[D \xi^{*}\left(Z_1,X\right)\right]}.
		$$
		In addition, note that $\mathbb{E}\left[\left. \xi^{*}\left(Z_1,X\right)\right|X\right] = 0$. Thus, by Lemma B.3 in \cite{kolesa2013estimation}, 
		\begin{equation}
			\label{expressioncml}
			\theta_{0} = \frac{\mathbb{E}\left[\omega(X)\tau\left(X\right)\right]}{\mathbb{E}\left[\omega(X)\right]},
		\end{equation}
		where 
		\begin{equation}
			\label{omega0}
			\omega(X) = \left(p_1 -p_0\right) \mathbb{E}\left[\left.\xi^{*}\left(Z_1,X\right)
			\right|p\left(Z_1,X\right)=p_1,X\right] \mathbb{P}\left(\left.p\left(Z_1,X\right) = p_1 \right| X\right).
		\end{equation}
		Next, observe that we can write 
		\begin{equation}
			\label{omega1}
			\mathbb{E}\left[\left.\xi^{*}\left(Z_1,X\right)
			\right|p\left(Z_1,X\right)=p_1,X\right] = p_1 - \mathbb{E}\left[\left.p\left(Z_1,X\right)\right|X\right].
		\end{equation}
		What is more, 
		\begin{equation}
			\label{omega2}
			\mathbb{E}\left[\left.p\left(Z_1,X\right)\right|X\right] = p_0 \mathbb{P}\left(\left.p\left(Z_1,X\right) = p_0 \right| X\right) + p_1 \mathbb{P}\left(\left.p\left(Z_1,X\right) = p_1 \right| X\right).
		\end{equation}
		Plugging \eqref{omega1} and \eqref{omega2} into \eqref{omega0} yields 
		\begin{equation*}
			\omega\left(X\right) = \left(p_1 - p_0\right)^2 \mathbb{P}\left(\left.p\left(Z_1,X\right) = p_0 \right| X\right) \mathbb{P}\left(\left.p\left(Z_1,X\right) = p_0 \right| X\right).
		\end{equation*}
		Now, note that 
		\begin{equation*}
			\begin{split}
				Var\left(\left.p(Z_1,X)\right|X\right) & = \mathbb{E}\left[\left.\left(p\left(Z_1,X\right) - \mathbb{E}\left[\left. p\left(Z_1,X\right) \right|X\right]\right)^2 \right| X\right] \\ & = \left(p_1 - \mathbb{E}\left[\left. p\left(Z_1,X\right) \right|X\right]\right)^2\mathbb{P}\left(\left.p = p_1\right|X\right) + \left(p_0 - \mathbb{E}\left[\left. p\left(Z_1,X\right) \right|X\right]\right)^2\mathbb{P}\left(\left.p = p_0\right|X\right) \\ & = \left(p_1 - p_0\right)^2 \mathbb{P}\left(\left.p\left(Z_1,X\right) = p_0 \right| X\right) \mathbb{P}\left(\left.p\left(Z_1,X\right) = p_0 \right| X\right),
			\end{split} 
		\end{equation*}
		where the last equality uses \eqref{omega2}. Thus, $\omega\left(X\right) = Var\left(\left.p(Z_1,X)\right|X\right)$. Plugging this result into \eqref{expressioncml} completes the proof. $\blacksquare$

		\section{Asymptotic Theory}
		\label{sasymptotictheory}
		The main result of this section indicates that inference on $\hat{\theta}_{CML}$ is standard, which means that standard errors can be computed as usual, albeit we are dealing with high-dimensional objects. This section exploits many of the theoretical devices that appear in \cite{chernozhukov2022locally}, with some modifications. The consistency of $\hat{\theta}_{CML}$ is shown in Section \ref{sconsistency}.

		Let $F_0$ be the distribution of the data, and let us define 
		\begin{equation*}
			\begin{split}
				g\left(Z,\theta,\eta,\kappa\right) & := \left(Y - \eta_{1}(X) - \theta\left(D - \eta_{2}(X)\right)\right)\kappa\left(Z_1,X\right) \\ \hat{g}_n(\theta_0) & := \frac{1}{n}\sum^L_{\ell = 1} \sum_{i \in I_\ell}\left(Y - \hat{\eta}_{1\ell}\left(X_i\right) - \theta_0\left(D_i - \hat{\eta}_{2\ell}\left(X_i\right)\right)\right) \hat{\kappa}_{\ell}\left(Z_{1i},X_i\right)  
			\end{split}
		\end{equation*}
		We start by imposing some standard regularity conditions for our model \eqref{eqmodel}.
		\begin{assumption}
			\label{amodel}
			(i) $\phi(X) \in L^2(X)$, (ii) $\theta_0$ uniquely satisfies \eqref{eqmodel}, (iii)  $\mathbb{E}\left[\left.\varepsilon^2\right|Z_1,X\right] < \infty $ a.s. (iv) $\xi^{*}(Z_1,X) \in L^2(Z_1,X)$, (v) $\mathcal{H}_\eta$ is a Hilbert space such that $ \mathcal{H}_\eta\subseteq L^2(X) \times L^2(X)$.
		\end{assumption}
		
		Moreover, we impose mean-square convergence conditions that involve the high-dimensional objects of the identifying moment \eqref{generallr}. 
		
		\begin{assumption}
			\label{aconvergencetozero}
			Let  $\int \hat{\eta}_{j\ell}(x)^2 F_0\left(dz\right) < \infty$, $j=1,2$, and $\int \hat{\kappa}_\ell(z_1,x)^2 F_0\left(dz\right) < \infty$, with probability approaching one. In addition, 
			\begin{itemize} 
				\item[(i)] $\int \left[\left(\eta_{01}(x) - \hat{\eta}_{1\ell}(x)\right) - \theta_0\left(\eta_{02}(x) - \hat{\eta}_{2\ell}(x)\right)\right]^2 \left[\hat{\kappa}_\ell\left(z_1,x\right) - \kappa_0\left(z_1,x\right)\right]^2 F_0(dz) \overset{p}{\to} 0$, 
				\item[(ii)] $\int \left[\left(\eta_{01}(x) - \hat{\eta}_{1\ell}(x)\right) - \theta_0\left(\eta_{02}(x) - \hat{\eta}_{2\ell}(x)\right)\right]^2 \kappa_0(z_1,x)^2 F_0(dz) \overset{p}{\to} 0$,
				\item[(iii)] $\int \left[Y - \eta_{01}(x) - \theta_0\left(d - \eta_{02}(x)\right)\right]^2 \left[\hat{\kappa}_\ell\left(z_1,x\right) - \kappa_0\left(z_1,x\right)\right]^2 F_0(dz) \overset{p}{\to} 0$.
			\end{itemize}
		\end{assumption}
		
		Also, we have to impose $\sqrt{n}-$convergence in the following sense
		\begin{assumption}
			\label{aratenorm}
			$ \sqrt{n} \left|\left| \hat{\eta}_{j\ell} - \eta_{0j} \right|\right|_2\left|\left| \hat{\kappa}_{\ell} - \kappa_{0} \right|\right|_2 \overset{p}{\to} 0$,  for $j = 1,2$. 
		\end{assumption}
		We next make the observation that, for all $\eta(X) \in \mathcal{H}_\eta$, and $\kappa(Z_1,X) \in L^2(Z_1,X)$,
		\begin{equation}
			\label{doublerobustness}
			\mathbb{E}\left[g\left(Z,\theta_0,\eta,\kappa_0\right)\right] = \mathbb{E}\left[g\left(Z,\theta_0,\eta_0,\kappa_0\right)\right] = \mathbb{E}\left[g\left(Z,\theta_0,\eta_0,\kappa\right)\right].  
		\end{equation}
		
		\begin{proposition}
			\label{pdobulerobustness}
			Let $\mathbb{E}\left[\left.\varepsilon^2\right|Z,X\right] < \infty$. Then, for all $\eta(X) \in \mathcal{H}_\eta \subseteq L^2(X) \times L^2(X)$, and $\kappa(Z_1,X) \in L^2(Z_1,X)$, \eqref{doublerobustness} holds.
		\end{proposition}

		\bigskip \noindent \textbf{Proof of Proposition \ref{pdobulerobustness}:} Observe that 
		\begin{equation}
			\label{dr1}
			\begin{split}
				\mathbb{E}\left[g\left(Z,\theta_0,\eta,\kappa_0\right)\right] & = \mathbb{E}\left[\left(Y 
				- \eta_1(X) - \theta_0\left(D - \eta_2(X)\right)\right)\kappa_0(Z_1,X)\right] \\ & = \mathbb{E}\left[\left(Y - \theta_0D - \phi(X)\right)\kappa_0(Z_1,X)\right] + \mathbb{E}\left[\left(\theta_0\eta_2(X) - \eta_1(X) + \phi(X)\right)\kappa_0(Z_1,X)\right] \\ & = \mathbb{E}\left[\left(Y - \theta_0D - \phi(X)\right)\kappa_0(Z_1,X)\right] \\ & = \mathbb{E}\left[g\left(Z,\theta_0,\eta_0,\kappa_0\right)\right].
			\end{split}
		\end{equation}
		where the third equality follows by the law of iterated expectations. Moreover, because of \eqref{eqmodel}, 
		$$
		0 = \mathbb{E}\left[g\left(W,\theta_0,\eta_0,\kappa_0\right)\right] = \mathbb{E}\left[g\left(W,\theta_0,\eta_0,\kappa\right)\right],
		$$
		as needed. $\blacksquare$

		\bigskip Proposition \ref{pdobulerobustness} implies that the property of robustness of a moment based on $g$, as characterized by \eqref{cond1}, is global. As we indicated above, global orthogonality also applies to $\kappa$. This feature is defined as double robustness by \cite{chernozhukov2022locally}. The fact that \eqref{doublerobustness} holds is important as it (along with cross-fitting) will make the conditions required for standard inference mild, and the proof of the result simpler. 
		
		The following lemma states the key result of this section. 
		\begin{lemma}
			\label{lkeyresult}
			Let Assumptions \ref{amodel} - \ref{aratenorm}. Then, 
			\begin{equation}
				\label{ekeyresusult}
				\sqrt{n}\hat{g}_n(\theta_0) = \frac{1}{n} \sum^n_{i=1} g\left(Z_i,\theta_0,\eta_0,\kappa_0\right) + o_p(1). 
			\end{equation}
		\end{lemma}
		
		\noindent \bigskip \textbf{Proof of Lemma \ref{lkeyresult}:} Let 
		$$
		m_1\left(Y,\theta,\eta\right) = Y - \eta_{1}(X) - \theta\left(D - \eta_{2}(X)\right).
		$$
		Furthermore, let us define
		\begin{equation*}
			\begin{split}
				\hat{R}_{1\ell i} & := \left[m_1\left(Y_i,\theta_0,\hat{\eta}_\ell\right) - m_1\left(Y_i,\theta_0,\eta_0\right)\right]\left[\hat{\kappa}_\ell(Z_{1i},X_i) - \kappa_0(Z_{1i},X_i)\right] \\ & = \left[\left(\eta_{01}(X_i) - \hat{\eta}_{1\ell}(X_i)\right) - \theta_0\left(\eta_{02}(X_i) - \hat{\eta}_{2\ell}(X_i)\right)\right]\left[\hat{\kappa}_\ell(Z_{1i},X_i) - \kappa_0(Z_{1i},X_i)\right], \\
				\hat{R}_{2\ell i} & :=  \left[m_1\left(Y_i,\theta_0,\hat{\eta}_\ell\right) - m_1\left(Y_i,\theta_0,\eta_0\right)\right]\kappa_0(Z_{1i},x_i) \\ & = \left[\left(\eta_{01}(X_i) - \hat{\eta}_{1\ell}(X_i)\right) - \theta_0\left(\eta_{02}(X_i) - \hat{\eta}_{2\ell}(X_i)\right)\right] \kappa_0(Z_{1i},X_i), \\ 
				\hat{R}_{3\ell i} & := m_1\left(Y_i,\theta_0,\eta_0\right) \left[\hat{\kappa}_\ell(Z_{1i},X_i) - \kappa_0(Z_{1i},X_i)\right] \\ & = \left[Y_i - \eta_{01}(X_i) - \theta_0\left(d - \eta_{02}(X_i)\right)\right] \left[\hat{\kappa}_\ell(Z_{1i},X_i) - \kappa_0(Z_{1i},X_i)\right]. 
			\end{split}
		\end{equation*}
		Then, we write 
		$$
		m\left(Y_i,\theta_0,\hat{\eta}_\ell\right)\hat{\kappa}_\ell(Z_{1i},X_i) - g\left(Z_i,\theta_0,\eta_0,\kappa_0\right) =  \hat{R}_{1\ell i} +  \hat{R}_{2\ell i} +  \hat{R}_{3\ell i}. 
		$$
		Let $\mathcal{Z}^c_\ell$ be observations that are not in $I_\mathcal{\ell}$. Then, $\hat{\eta}_{1\ell}$, $\hat{\eta}_{2\ell}$, $\hat{\kappa}_{\ell}$ depend only on  $\mathcal{Z}^c_\ell$. Next, observe that 
		
		\begin{equation*}
			\begin{split}
				\mathbb{E}\left[\left.\hat{R}_{1\ell i} + \hat{R}_{2\ell i}\right|\mathcal{Z}^c_\ell \right] & = \mathbb{E}\left[\left.\left(Y - \hat{\eta}_{1\ell}(X) - \theta_0\left(D - \hat{\eta}_{2\ell}(X)\right)\right)\kappa_0(Z_1,X)\right|\mathcal{Z}^c_\ell\right] \\ & + \mathbb{E}\left[\left.\left(Y - \hat{\eta}_{1\ell}(X) - \theta_0\left(D - \hat{\eta}_{2\ell}(X)\right)\right)\left(\hat{\kappa}_\ell\left(Z_1,X\right)-\kappa_0(Z_1,X)\right)\right|\mathcal{Z}^c_\ell\right]. \\ \mathbb{E}\left[\left.\hat{R}_{3\ell i} \right|\mathcal{Z}^c_\ell \right] & = 0.
			\end{split}
		\end{equation*}
		Note that 
		\begin{align}
			\left|\frac{1}{\sqrt{n}} \sum_{i \in I_\ell} \mathbb{E}\left[\left. \hat{R}_{1\ell i} + \hat{R}_{2\ell i} + \hat{R}_{3\ell i}\right|\mathcal{Z}^c_\ell\right]\right| & \leq \sqrt{n} \left|\mathbb{E}\left[\left. \hat{R}_{1\ell i} + \hat{R}_{2\ell i} \right|\mathcal{Z}^c_\ell\right]\right|\nonumber \\ & \leq \sqrt{n} \left| \mathbb{E}\left[\left.\left(Y - \hat{\eta}_{1\ell}(X) - \theta_0\left(D - \hat{\eta}_{2\ell}(X)\right)\right)\kappa_0(Z_1,X)\right|\mathcal{Z}^c_\ell\right]\right|  \label{sqrtn1}\\ & + \sqrt{n}\left| \mathbb{E}\left[\left.\left(Y - \hat{\eta}_{1\ell}(X) - \theta_0\left(D - \hat{\eta}_{2\ell}(X)\right)\right)\left(\hat{\kappa}_\ell\left(Z_1,X\right)-\kappa_0(Z_1,X)\right)\right|\mathcal{Z}^c_\ell\right] \right|. \label{sqrtn2}
		\end{align}
		By  the Cauchy-Schwarz inequality, Assumption \ref{aratenorm} implies that  \eqref{sqrtn2} converges in probability to zero. Moreover, the result in \eqref{doublerobustness} indicates that \eqref{sqrtn1} is zero, with probability approaching one. Therefore, 
		\begin{equation}
			\label{ers}
			\left|\frac{1}{\sqrt{n}} \sum_{i \in I_\ell} \mathbb{E}\left[\left. \hat{R}_{1\ell i} + \hat{R}_{2\ell i} + \hat{R}_{3\ell i}\right|\mathcal{Z}^c_\ell\right]\right| \overset{p}{\to} 0.
		\end{equation}
		Next, as conditional on $\mathcal{Z}^c_\ell$, observations are mutually independent, we have 
		\begin{equation*}
			\begin{split}
				\mathbb{E}\left[\left. \left\{ \hat{R}_{k\ell i} - \mathbb{E}\left[\left.\hat{R}_{k\ell i\ell}\right|\mathcal{Z}^c_\ell\right]\right\}^2 \right|\mathcal{Z}^c_\ell\right] & = \frac{n_\ell}{n}Var\left(\left. \hat{R}_{k\ell i\ell}\right|\mathcal{Z}^c_\ell\right) \\&  \leq Var\left(\left. \hat{R}_{k\ell i\ell}\right|\mathcal{Z}^c_\ell\right) \\ & \leq \mathbb{E}\left[\left. \hat{R}^2_{k\ell i\ell}\right|\mathcal{Z}^c_\ell\right] \\ & \overset{p}{\to} 0,
			\end{split}
		\end{equation*}
		where the last display follows from Assumption \ref{aconvergencetozero}. Hence, by the triangle inequality and the conditional Markov's inequality, we obtain
		\begin{equation}
			\label{rexpzero}
			\frac{1}{\sqrt{n}} \sum_{i \in I_\ell} \left(\hat{R}_{1 \ell i} + \hat{R}_{2 \ell i} + \hat{R}_{3 \ell i}  - \mathbb{E}\left[\left. (\hat{R}_{1 \ell i} + \hat{R}_{2 \ell i} + \hat{R}_{3 \ell i}\right| \mathcal{Z}^c_\ell\right]\right) \overset{p}{\to} 0.
		\end{equation}
		But because of \eqref{ers}, \eqref{rexpzero} actually implies 
		$$
		\frac{1}{\sqrt{n}} \sum_{i \in I_\ell} \left(\hat{R}_{1 \ell i} + \hat{R}_{2 \ell i} + \hat{R}_{3 \ell i} \right) \overset{p}{\to} 0,
		$$
		which leads to conclude that \eqref{ekeyresusult} holds. $\blacksquare$
		
		\bigskip Let $\tilde{\theta}_\ell$ be a consistent estimator of $\theta_0$ using observations not in $I_\ell$, possibly based on a non-LR moment.  We then impose additional conditions.
		\begin{assumption}
			\label{aextra}
			$\int \left(g\left(z,\tilde{\theta}_\ell, \hat{\eta}_\ell, \hat{\kappa}_\ell\right) - g\left(z,\theta_0, \hat{\eta}_\ell, \hat{\kappa}_\ell\right)\right)^2 F_0\left(dz\right) \overset{p}{\to} 0$. 
		\end{assumption}

		\begin{assumption}
			\label{aextr2}
			$\int \left|\eta_{02}(x)\kappa_0(z_1,x) - \hat{\eta}_{2\ell}(x)\hat{\kappa}_{\ell}(z_1,x)\right| F_0(dw) \overset{p}{\to} 0$.
		\end{assumption}
		
		The following theorem guarantees that inference on $\hat{\theta}_{CML}$ is standard. 
		\begin{theorem}
			\label{tstandardinference}
			Let Assumptions \ref{amodel} - \ref{aextr2} hold. In addition, assume $\mathbb{E}\left[\left. D \right|Z_1,X\right] \neq \mathbb{E}\left[\left. D \right|X\right]$, and that $\hat{\theta}_{CML} \overset{p}{\to} \theta_0$. Then, 
			$$
			\sqrt{n}\left(\hat{\theta}_{CML} - \theta_0\right) \overset{d}{\to} N\left(0, \mathcal{V}\right), \;\;\;\; \mathcal{V} = \frac{\mathbb{E}\left[\varepsilon^2 \xi^{*}\left(Z_1,X\right)^2
				\right]}{\mathbb{E}\left[\xi^{*}\left(Z_1,X\right)
				\right]^2}. 
			$$
			Additionally, a consistent estimator of $\mathcal{V}$ is 
			$$
			\hat{V} = \frac{\frac{1}{n}\sum^L_{\ell = 1} \sum_{i \in I_\ell} \left(Y_{i} - \hat{\eta}_{1\ell}(X_i) - \tilde{\theta}_\ell\left(D_i - \hat{\eta}_{2\ell}\right)\right)^2 \hat{\kappa}_\ell(Z_{1i},X_i)^2}{\left(\frac{1}{n}\sum^L_{\ell = 1}  \sum_{i \in I_\ell}  \hat{\kappa}_\ell(Z_{1i},X_i)\right)^2}.
			$$
		\end{theorem}
		
		\bigskip To prove Theorem \ref{tstandardinference} we will use Lemma \ref{lkeyresult} along with two additional lemmas that we now state. 
		
		Let us define 
		\begin{equation*}
			\begin{split}
				\Psi & := \mathbb{E}\left[\varepsilon^2 \kappa_0\left(Z_1,X\right)^2
				\right], \\ 
				\hat{\Psi}  & := \frac{1}{n}\sum^L_{\ell = 1} \sum_{i \in I_\ell} \left(Y_{i} - \hat{\eta}_{1\ell}(X_i) - \tilde{\theta}_\ell\left(D_i - \hat{\eta}_{2\ell}\right)\right)^2 \hat{\kappa}_\ell(Z_{1i},X_i)^2. 
			\end{split}
		\end{equation*}
		
		\begin{lemma}
			\label{lsecond}
			Let Assumption \ref{aextra} hold. Then, $\hat{\Psi} \overset{p}{\to} \Psi$.
		\end{lemma}
		
		\noindent \bigskip \textbf{Proof of Lemma \ref{lsecond}:} Let $g_i = \psi\left(Z_i, \theta_0,\eta_0,\kappa_0\right)$ and $\tilde{\Psi} = \frac{1}{n} \sum^n_{i=1} g^2_i$. In addition, let $\hat{R}_{k\ell i}$, $k = 1,2,3$, be defined as in the proof of Lemma \ref{lkeyresult}, and define now 
		$$
		\hat{R}_{4 \ell i} : = g\left(z_i,\tilde{\theta}_\ell, \hat{\eta}_\ell,\hat{\kappa}_\ell\right) - g\left(z_i,\theta_0, \hat{\eta}_\ell,\hat{\kappa}_\ell\right).
		$$
		Moreover, let $\hat{g}_{i \ell}: = \left(Y_i - \hat{\eta}_{1\ell}(X_i) - \tilde{\theta}_\ell \left(D_i - \hat{\eta}_{2\ell}(X_i)\right)\right) \hat{\kappa}_\ell\left(Z_{1i},X_i\right)$, and observe that 
		$$
		\hat{g}_{i \ell} - g_i = \hat{R}_{1 \ell i} + \hat{R}_{2 \ell i} + \hat{R}_{3 \ell i} + \hat{R}_{4 \ell i}. 
		$$
		By Assumption \ref{aextra}, we have that $\mathbb{E}\left[\left. \hat{R}^2_{4\ell i} \right| \mathcal{Z}^c_\ell\right] \overset{p}{\to} 0$. What is more, by Assumption \ref{aconvergencetozero}, $\mathbb{E}\left[\left. \hat{R}^2_{k\ell i} \right| \mathcal{Z}^c_\ell\right] \overset{p}{\to} 0$, $k = 1,2,3$. Using this, we can write 
		\begin{equation*}
			\begin{split}
				\mathbb{E}\left[\left. \frac{1}{n} \sum_{i \in I_\ell} \left| \hat{g}_{i \ell} - g_i \right|^2 \right| \mathcal{Z}^c_\ell\right] & = \frac{n_\ell}{n} \mathbb{E}\left[\left.\left|\hat{g}_{i \ell} - g_i\right|^2\right| \mathcal{Z}^c_\ell\right] \leq \frac{C n_\ell}{n} \sum^4_{k=1} \mathbb{E}\left[\left.\left|\hat{R}_{k \ell i} \right|^2 \right| \mathcal{W}^c_{\ell}\right] \overset{p}{\to} 0.
			\end{split}
		\end{equation*}
		Therefore, by the conditional Markov's inequality, $\frac{1}{n} \sum_{i \in I_\ell} \left| \hat{g}_{i \ell} - g_i \right|^2 \overset{p}{\to} 0$, for each $\ell = 1,\cdots, L$. Then, using the triangle inequality and the Cauchy-Schwartz inequality, 
		\begin{equation}
			\begin{split}
				\left|\hat{\Psi} - \tilde{\Psi}  \right| & \leq  \sum^L_{\ell = 1}\left(\frac{1}{n} \sum_{i \in I_\ell} \left|\hat{g}_{i \ell} - g_i \right|^2 + 2 \frac{1}{n}\sum_{i \in I_\ell} \left|\hat{g}_{i \ell} - g_i \right| \left|g_i\right| \right) \\ & \leq \sum^L_{\ell = 1}\left(\frac{1}{n} \sum_{i \in I_\ell} \left|\hat{g}_{i \ell} - g_i \right|^2 + 2 \sqrt{\frac{1}{n}\sum_{i \in I_\ell} \left|\hat{g}_{i \ell} - g_i \right|^2} \sqrt{\frac{1}{n} \sum_{i \in I_\ell} \left|g_i\right|^2} \right) \\ & = o_p(1)\left(1 + O_p(1)\right)  \overset{p}{\to} 0. 
			\end{split}
		\end{equation}
		By the law of large numbers, we know $\tilde{\Psi} \overset{p}{\to} \Psi$. Therefore, by the triangle inequality, $\hat{\Psi} \overset{p}{\to} \Psi$, as needed. $\blacksquare$
		
		\begin{lemma}
			\label{thirdlemma}
			Let Assumption \ref{aextr2} hold.  Then, 
			\begin{equation}
				\label{ethirdlemma}
				\frac{1}{n} \sum^L_{\ell = 1} \sum_{i \in I_\ell} \left(D_i - \hat{\eta}_{2\ell}\left(X_i\right)\right) \hat{\kappa}_{\ell}\left(Z_{1i},X_i\right) \overset{p}{\to} \mathbb{E}\left[\left(D - \eta_{02}\left(X\right)\right) \kappa_0\left(Z_1,X\right)\right]. 
			\end{equation}
		\end{lemma}
		
		\bigskip \noindent \textbf{Proof of Lemma \ref{ethirdlemma}:} Let 
		\begin{equation*}
			\begin{split}
				\hat{\Upsilon}_\ell & = \frac{1}{n} \sum_{i \in I_\ell} \left(D_i - \hat{\eta}_{2\ell}\left(X_i\right)\right) \hat{\kappa}_{\ell}\left(Z_{1i},X_i\right) \\  \tilde{\Upsilon}_{\ell} & = \frac{1}{n_\ell} \sum_{i \in I_\ell} \left(D_i - \eta_{02}\left(X_i\right)\right) \kappa_0\left(Z_{1i},X_i\right) \\ \Upsilon & = \mathbb{E}\left[\left(D - \eta_{02}\left(X\right)\right) \kappa_0\left(Z_1,X\right)\right].
			\end{split}
		\end{equation*}
		Note that, by the triangle inequality,  
		$$
		\mathbb{E}\left[\left.\left|\hat{\Upsilon}_\ell - \tilde{\Upsilon}_\ell \right| \right| \mathcal{W}^c_{\ell}\right] \leq  \mathbb{E}\left[\left. \left|\left(\eta_{02}\left(X_i\right)\kappa_0\left(Z_{1i},X_i\right) - \hat{\eta}_{2\ell}\left(X_i\right)\hat{\kappa}_\ell\left(Z_{1i},X_i\right)\right)\right| \right| \mathcal{Z}^c_\ell \right] \overset{p}{\to} 0,
		$$
		which follows by Assumption \ref{aextr2}. Then, by the conditional Markov's inequality, $\hat{\Upsilon}_\ell \overset{p}{\to} \tilde{\Upsilon}_\ell$. Furthermore, by the law of large numbers, $\tilde{\Upsilon}_\ell \overset{p}{\to} \Upsilon$. Therefore, the result in \eqref{ethirdlemma} holds. $\blacksquare$
		
		\bigskip \noindent \textbf{Proof of Theorem \ref{tstandardinference}:} The result in the theorem follows by Lemmas \ref{lkeyresult}-\ref{thirdlemma} and the Central Limit Theorem, using standard arguments, as in, e.g., the proof of Proposition 21.20 in \cite{ruud2000introduction}. $\blacksquare$
		
		\section{Consistency}
		\label{sconsistency}
		This section exhibits the conditions required for the consistency of $\hat{\theta}_{CML}$, and its proof. 
		
		\begin{theorem}
			\label{tconsistency}
			Let the following assumptions hold: (i) Assumption \ref{amodel} (ii); (ii) $\mathbb{E}\left[\left. D \right|Z_1,X\right] \neq \mathbb{E}\left[\left. D \right|X\right]$; (iii) $\int \left|\eta_{01}(x)\kappa_0(z_1,x) - \hat{\eta}_{1\ell}(x)\kappa_{\ell}(z_1,x)\right|F_0\left(dz\right) \overset{p}{\to} 0$; (iv) Assumption \ref{aextr2}.  Then, $\hat{\theta}_{CML} \overset{p}{\to} \theta_0$.	
		\end{theorem}
		
		\bigskip \noindent \textbf{Proof of Theorem \ref{tconsistency}:} Let 
		
		\begin{equation*}
			\begin{split}
				\hat{M}_{1\ell} & = \frac{1}{n_\ell} \sum_{i \in I_\ell} \left(Y_{i} - \hat{\eta}_{1\ell}(X_i)\right) \hat{\kappa}_\ell\left(Z_{1i},X_i\right), \\ 
				\tilde{M}_{1\ell} & = \frac{1}{n_\ell} \sum_{i \in I_\ell} \left(Y_{i} - \eta_{01}(X_i)\right) \kappa_0\left(Z_{1i},X_i\right), \\
				\hat{M}_{2\ell} & = \frac{1}{n_\ell} \sum_{i \in I_\ell} \left(D_i - \hat{\eta}_{2\ell}(X_i)\right) \hat{\kappa}_\ell\left(Z_{1i},X_i\right), \\ 
				\tilde{M}_{2\ell} & = \frac{1}{n_\ell} \sum_{i \in I_\ell} \left(Y_{i} - \eta_{02}(X_i)\right) \kappa_0\left(Z_{1i},X_i\right).
			\end{split}
		\end{equation*}
		Next, observe that, by  \textit{(iii)}, 
		\begin{equation*}
			\begin{split}
				\mathbb{E}\left[\left. \left|\hat{M}_{1\ell} - \tilde{M}_{1\ell} \right|\right|\mathcal{Z}^c_\ell\right] & \leq \mathbb{E}\left[\left. \left|\eta_{01}\left(X\right)\kappa_{0}\left(Z,X\right) - \hat{\eta}_{1\ell}\left(X\right)\hat{\kappa}_\ell\left(Z_1,X\right) \right| \right|\mathcal{Z}^c_{\ell}\right] \\ & \overset{p}{\to} 0.
			\end{split}
		\end{equation*}
		Then, by the conditional Markov's inequality, $\hat{M}_{1\ell} \overset{p}{\to} \tilde{M}_{1\ell}$. In addition, by the law of large numbers, $\tilde{M}_{1\ell} \overset{p}{\to} \mathbb{E}\left[\left(Y - \eta_{01}\left(X\right)\right)\kappa_0\left(Z_1,X\right)\right]$. Hence, by the triangle inequality, 
		\begin{equation}
			\label{econsistency1}
			\hat{M}_{1\ell} \overset{p}{\to} \mathbb{E}\left[\left(Y - \eta_{01}\left(X\right)\right)\kappa_0\left(Z_1,X\right)\right].
		\end{equation}
		By the same token as before and \textit{(iv)}, we can conclude that 
		\begin{equation}
			\label{econsistency2}
			\hat{M}_{2\ell} \overset{p}{\to} \mathbb{E}\left[\left(D - \eta_{02}\left(X\right)\right)\kappa_0\left(Z_1,X\right)\right].
		\end{equation}
		Therefore, by the results in \eqref{econsistency1}-\eqref{econsistency2}, and the continuous mapping theorem, we can show $\hat{\theta}_{CML} \overset{p}{\to} \theta_0$, as needed. $\blacksquare$
		
		\section{Additional computations for the Monte Carlo experiment}
		\label{sadditionalmc}
		In the Monte Carlo experiment, we consider three estimands: 
		\begin{equation*}
			\begin{split}
				\tau_{LATE} & = \frac{\mathbb{E}\left[\pi\left(X_1\right) \tau\left(X_1\right)\right]}{\mathbb{E}\left[\pi\left(X_1\right)\right]},  \\ 
				\theta_{0} & = \frac{\mathbb{E}\left[\pi\left(X_1\right)^2 Var\left(\left. Z_1\right|X_1\right) \tau\left(X_1\right)\right]}{\mathbb{E}\left[\pi\left(X_1\right)^2 Var\left(\left. Z_1\right|X_1\right)\right]},\\ 
				\theta_{DML} & = \frac{\mathbb{E}\left[c(X_1) \pi(X_1) Var\left(Z_1|X_1\right) \tau(X_1)\right]}{\mathbb{E}\left[c(X_1) \pi(X_1) Var\left(Z_1|X_1\right)\right]}.
			\end{split}
		\end{equation*}
		We next compute each of the terms involved in the previous expressions. 
		If $X_1 = \bm{1}\left(\delta_i \geq 0\right)$, we have that 
		\begin{equation*}
			\begin{split}
				\pi\left(X_1 = 1\right) & = \mathbb{P}\left(\left.D(1) > D(0)\right|X_1 = 1\right) \\ 
				& = \mathbb{P}\left(\left.\Phi^{-1}\left(s_2\right) \leq \delta \leq \Phi^{-1}\left(1-s_1\right)\right|X_1 = 1 \right) \\ 
				& = \mathbb{P}\left(\left.\Phi^{-1}\left(s_2\right) \leq \delta \leq \Phi^{-1}\left(1-s_1\right)\right|\delta \geq 0\right) \\ 
				& = \mathbb{P}\left(\left. \delta \leq \Phi^{-1}\left(1-s_1\right)\right|\delta \geq 0\right) - \mathbb{P}\left(\left. \delta \leq \Phi^{-1}\left(s_2\right)\right|\delta \geq 0\right) \\ 
				& = \frac{\Phi\left(\Phi^{-1}\left(1-s_1\right)\right) - \Phi\left(0\right)}{ 1 - \Phi\left(0\right)} - \frac{\Phi\left(\Phi^{-1}\left(s_2\right)\right) - \Phi\left(0\right)}{ 1 - \Phi\left(0\right)} \\ 
				& = \frac{1-s_1 - s_2}{ 1- \Phi\left(0\right)},
			\end{split}
		\end{equation*}
		provided that $s_1$ and $s_2$ are such that $0 \leq \pi\left(X_1 = 1\right) \leq 1$. By the same toke as before, we can show 
		\begin{equation}
			\begin{split}
				\pi\left(X_1 = 0\right) & = \mathbb{P}\left(\left.D(1) < D(0)\right|X_1 = 0\right)\\
				& = \frac{1-s_1 - s_2}{\Phi\left(0\right)},
			\end{split}
		\end{equation}
		provided that $s_1$ and $s_2$ are such that $0 \leq \pi\left(X_1 = 0\right) \leq 1$. If $X_1 = \bm{1}\left(\delta \geq -\infty\right)$, defiers are completely rule out, and
		$\pi(X) = 1 - s_1 - s_2$. 
		
		\begin{equation*}
			\begin{split}
				\tau\left(X_1\right) & = \mathbb{E}\left[\left.Y(1) - Y(0)\right|D(1) \neq D(0), X_1\right] \\ 
				& = \mathbb{E}\left[\left.\tau\right| D(1) \neq D(0), X \right] \\ 
				& = \mathbb{E}\left[\left.\tau\right|X_1\right] \\ 
				& = \mathbb{E}\left[\left. \mathbb{E}\left[\left. \tau \right|\delta, X_1 \right] \right|X_1\right] \\ 
				& = \mathbb{E}\left[\left. \mathbb{E}\left[\left. \tau \right|\delta\right] \right|X_1\right] \\ 
				& = \rho_{\delta \tau}\sigma_\tau \mathbb{E}\left[\left. \delta \right| X_1\right],
			\end{split}
		\end{equation*}
		where the first, the second, and fifth equalities follow by definition, the third equality uses that $\tau$ is independently distributed from $D(1)$ and $D(0)$, the fourth equality employs the law of iterated expectations, and the last one comes from the fact that $\mathbb{E}\left[\left. \tau \right| \delta\right] = \rho_{\delta \tau} \sigma_\tau \delta$. Next, as $X_1 = \bm{1}\left(\delta \geq 0\right)$, we should write $\mathbb{E}\left[\left. \delta \right| X_1\right] = \mathbb{E}\left[\left. \delta \right| X_1=0\right] + X_1 \left(\mathbb{E}\left[\left. \delta \right| X_1 = 1\right] - \mathbb{E}\left[\left. \delta \right| X_1 = 0\right]\right) = \mathbb{E}\left[\left. \delta \right| \delta < 0\right] + X_1 \left(\mathbb{E}\left[\left. \delta \right| \delta \geq 0\right] - \mathbb{E}\left[\left. \delta \right| \delta < 0\right]\right)$. This implies that, 
		$$
		\tau\left(X_1\right) = \rho_{\delta \tau} \sigma_\tau \left[-\frac{\phi\left(0\right)}{\Phi\left(0\right)} + 2X_1 \frac{\phi\left(0\right)}{\Phi\left(0\right)}\right],
		$$
		where $\phi$ is the standard normal pdf. What is more, when $X_1 = \bm{1}\left(\delta > -\infty\right)$, 
		$$
		\tau(X_1) = \rho_{\delta \tau}\sigma_\tau \mathbb{E}\left[\delta\right] = 0.
		$$
		Recall that $Z_1 \in \left\{0,1 \right\}$, and in particular, $\mathbb{P}\left(\left.Z_1 = 1\right|X_1\right) = \Phi\left(\alpha_z + \beta_{XZ} X_1\right)$, thus, 
		$$
		Var\left(\left. Z_1 \right|X_1\right) = \Phi\left(\alpha_z + \beta_{XZ} X_1\right)\left(1-\Phi\left(\alpha_z + \beta_{XZ} X_1\right)\right). 
		$$
		Moreover, if $Z_1$ is independent of $X_1$, we let $\beta_{XZ} = 0$, and the previous variance is constant. $c(X_1)$ can be simply written as follows
		$$
		c(X_1) = -1 + 2X_1.
		$$
		
		Therefore, if $X_1 = \bm{1}\left(\delta \geq 0\right)$, 
		\begin{equation*}
			\begin{split}
				\tau_{LATE} & = \frac{1}{2}\rho_{\delta \tau} \sigma_\tau \frac{\phi(0)}{\Phi(0)} \left(\frac{1-s_1-s_2}{1-\Phi(0)} - \frac{1-s_1-s_2}{\Phi(0)}\right) = 0, \\
				\theta_0 & = \frac{\frac{1}{2} \left(\frac{1-s_1-s_2}{1-\Phi(0)}\right)^2 \rho_{\delta \tau} \sigma_{\tau} \frac{\phi(0)}{\Phi(0)}\left(\Phi\left(\alpha_Z + \beta_{XZ}\right)\left(1 - \Phi\left(\alpha_Z + \beta_{XZ}\right)\right) - \Phi\left(\alpha_Z \right)\left(1 - \Phi\left(\alpha_Z \right)\right)\right)}{\frac{1}{2}\left(\frac{1-s_1-s_2}{1-\Phi(0)}\right)^2\left(\Phi\left(\alpha_Z + \beta_{XZ}\right)\left(1 - \Phi\left(\alpha_Z + \beta_{XZ}\right)\right) + \Phi\left(\alpha_Z \right)\left(1 - \Phi\left(\alpha_Z \right)\right)\right)}, \\ 
				\theta_{DML} & =\frac{\frac{1}{2} \frac{1-s_1-s_2}{1-\Phi(0)} \rho_{\delta \tau} \sigma_{\tau} \frac{\phi(0)}{\Phi(0)}\left(\Phi\left(\alpha_Z + \beta_{XZ}\right)\left(1 - \Phi\left(\alpha_Z + \beta_{XZ}\right)\right) + \Phi\left(\alpha_Z \right)\left(1 - \Phi\left(\alpha_Z \right)\right)\right)}{\frac{1}{2}\frac{1-s_1-s_2}{1-\Phi(0)}\left(\Phi\left(\alpha_Z + \beta_{XZ}\right)\left(1 - \Phi\left(\alpha_Z + \beta_{XZ}\right)\right) -\Phi\left(\alpha_Z \right)\left(1 - \Phi\left(\alpha_Z \right)\right)\right)}. 
			\end{split}
		\end{equation*}
		What is more, observe that when $X_1 = \bm{1}\left(\delta > -\infty\right)$, $\tau(X_1) = 0$, and thus $\tau_{LATE} = \theta_0 = \theta_{DML}= 0$. 
	\end{appendix}

	\clearpage
	\phantomsection
	\bibliographystyle{ectabib}
	\bibliography{ComplianceCMR}
	
\end{document}